\documentclass[times,twocolumn,final]{elsarticle}

\usepackage{jasr}
\usepackage{framed,multirow}

\usepackage{amssymb}
\usepackage{latexsym}

\usepackage[switch]{lineno}

\usepackage{url}
\usepackage{xcolor}
\definecolor{newcolor}{rgb}{.8,.349,.1}
\usepackage[normalem]{ulem}

\usepackage[citebordercolor=white]{hyperref}

\usepackage{amsmath}		
\usepackage{comment}
\usepackage[normalem]{ulem}

\newif\ifhighlight
\highlightfalse

\journal{Advances in Space Research}

\begin{document}

\verso{Igor I. Baliukin \textit{et al.}}

\begin{frontmatter}

\title{Understanding the IBEX ribbon using the kinetic model of pickup proton transport in a scatter-free limit}

\author[1,2,3]{Igor I. \snm{Baliukin}\corref{cor1}}
\cortext[cor1]{Corresponding author: email address: igor.baliukin@cosmos.ru}

\author[1,4,5]{Vladislav V. \snm{Izmodenov}}

\affiliation[1]{organization={Space Research Institute of Russian Academy of Sciences},
                addressline={Profsoyuznaya Str. 84/32},
                city={Moscow},
                postcode={117997},
                country={Russia}}
                
\affiliation[2]{organization={HSE University},
                addressline={20 Myasnitskaya Ulitsa},
                city={Moscow},
                postcode={101000},
                country={Russia}}
                
\affiliation[3]{organization={State Key Laboratory of Solar Activity and Space Weather, National Space Science Center, Chinese Academy of Sciences},
                city={Beijing},
                postcode={100190},
                country={China}}
                
\affiliation[4]{organization={Lomonosov Moscow State University, Moscow Center for Fundamental and Applied Mathematics},
                addressline={GSP-1, Leninskie Gory},
                city={Moscow},
                postcode={119991},
                country={Russia}}
                
\affiliation[5]{organization={Institute for Problems in Mechanics},
                addressline={Vernadskogo 101-1},
                city={Moscow},
                postcode={119526},
                country={Russia}}

\received{25 December 2024}
\finalform{5 May 2025}
\accepted{9 May 2025}

\begin{abstract}

One of the most remarkable discoveries by the \textit{Interstellar Boundary Explorer (IBEX)} is the ribbon -- a narrow band of enhanced energetic neutral atom (ENA) fluxes observed in the sky. The prevailing explanation attributes the \textit{IBEX} ribbon to the secondary ENA mechanism. In this process, ``primary'' hydrogen ENAs, produced via charge exchange between solar wind protons and interstellar hydrogen atoms within the heliosphere, travel beyond the heliopause and undergo further charge exchange with protons of the local interstellar medium (LISM), generating pickup protons. Some of these pickup protons subsequently experience charge exchange with interstellar hydrogen atoms, forming ``secondary'' ENAs, some of which travel back toward the Sun and are detected by the \textit{IBEX}.

This paper presents a kinetic model developed to simulate secondary ENA fluxes. Ribbon simulations are performed using global distributions of plasma and hydrogen atoms in the heliosphere derived from a kinetic-magnetohydrodynamic model of the solar wind interaction with the LISM. The model accounts for all relevant primary ENA populations, including neutralized thermal solar wind protons, neutralized pickup protons, and ENAs originating in the inner heliosheath. The transport of pickup protons beyond the heliopause is described by the focused transport equation for a gyrotropic velocity distribution in the scatter-free limit, assuming no pitch-angle scattering or energy diffusion.

Our simulations qualitatively reproduce \textit{IBEX-Hi} (0.5--6 keV) ribbon observations and exhibit good quantitative agreement at low heliolatitudes. However, the model underestimates fluxes at high heliolatitudes, likely due to the omission of non-stationary solar wind behavior in the stationary framework used in this work. The study highlights the importance of ENAs from the inner heliosheath, a population considered for the ribbon production in the frame of the kinetic model of pickup proton transport in the heliosphere for the first time, for accurately reproducing ribbon fluxes observed by \textit{IBEX-Hi} at the highest energy steps.

\end{abstract}




\end{frontmatter}


\section{Introduction} \label{sec:intro}

The interaction between the solar wind (SW) and the local interstellar medium (LISM), which is partially ionized plasma composed mainly of protons and hydrogen atoms, forms the heliosphere, a protective bubble surrounding our solar system that is moving through the LISM with a bulk velocity of $\sim$26 km/s \citep[e.g.][]{witte2004, mccomas2015}. The solar wind expansion eventually leads to the formation of the termination shock (TS), where the solar wind is slowed down and heated. The heliopause (HP) marks the boundary between the magnetized solar wind plasma and the charged component of the LISM, with its embedded interstellar magnetic field (IsMF). The inner heliosheath (IHS) is the region between the termination shock and the heliopause, while the region of compressed LISM plasma between the heliopause and a bow shock (BS), where the interstellar flow is slowed down, was traditionally termed the outer heliosheath (OHS). However, recent studies suggest that a broad bow wave, rather than a bow shock, is formed in the LISM flow upstream of the heliopause \citep[see discussions in, e.g.][]{izmod2009, mccomas2012a, zank2013}. Throughout this paper, we will use the term ``the outer heliosheath'' to indicate the part of the LISM affected by the presence of the heliosphere (note that some authors prefer to use the term ``very local interstellar medium'', or VLISM, to denote this region of space).

The \textit{Interstellar Boundary Explorer (IBEX)} mission, launched in 2008, has significantly advanced our understanding of SW/LISM interaction by observing energetic neutral atoms (ENAs) originating at the heliospheric boundaries \citep{mccomas2009}. IBEX carries two ENA cameras: IBEX-Lo, which measures ENAs with energies from 10 eV to 2 keV \citep{fuselier2009}, and IBEX-Hi, which measures ENAs from 0.5 to 6 keV \citep{funsten2009}. These instruments provide all-sky maps of ENA fluxes, revealing the structure and dynamics of the heliospheric interface.

One of IBEX's most significant discoveries is the so-called IBEX ribbon -- a narrow, nearly circular band of enhanced ENA fluxes in the sky, which is superimposed on top of a more diffuse flux background \citep{mccomas2009_science}. This unexpected discovery required a new understanding of the SW/LISM interaction. The ribbon geometry appeared to be ordered by the interstellar magnetic field, specifically aligning with lines of sight where the radial component of the magnetic field is close to zero. Originally motivated by the differences in the interstellar H and He flow directions \citep{lallement2005} and the asymmetry in the Voyager 1 and 2 crossings of the termination shock \citep{stone2008}, this magnetic field direction was incorporated into global heliospheric models \citep{izmod2005b, opher2007, pogorelov2009}. In these models, the observed location of the ribbon aligned close to perpendicular to the magnetic field \citep{schwadron2009}. There is also the latitudinal ordering of ribbon ENAs by energy: the higher-energy ribbon fluxes are observed primarily at high ecliptic latitudes and low-energy fluxes -- at low latitudes. This behavior reflects the solar wind structure during the solar minimum \citep{mccomas2012b}.

Multiple hypotheses have been proposed to explain the formation of the IBEX ribbon \citep{mccomas2014}. The commonly accepted hypothesis for the ribbon formation is the secondary ENA mechanism. It proposes that ``primary'' hydrogen ENAs, generated by charge exchange between the solar wind protons and interstellar H atoms inside the heliosphere, propagate outward and cross the heliopause. Once beyond the heliopause, these primary ENAs undergo a second charge exchange with the LISM protons, creating hydrogen pickup ions (PUIs). Some of these pickup protons subsequently experience a third charge exchange interaction with interstellar hydrogen atoms, becoming ``secondary'' ENAs that can travel back toward the Sun and be detected by IBEX.

\citet{heerikhuisen2016} and \citet{gamayunov2017} noted that different primary ENA sources should be considered in the ribbon analysis, namely, the H atoms generated in the supersonic solar wind and inner heliosheath regions.
\citet{schwadron2019} proposed a unifying explanation for the ribbon, suggesting it originates from a combination of three primary ENA sources: (1) the neutral solar wind (NSW) population, which originates from the charge exchange of solar wind thermal protons with interstellar H atoms inside the TS, (2) neutralized pickup ions (NPIs) from the supersonic solar wind (SSW) region, (3) ENAs from the inner heliosheath. They concluded that the secondary suprathermal atoms from the inner heliosheath may play a crucial role in shaping the ribbon structure, particularly at higher energy levels, leading to the broadening of the ribbon.

Several numerical models have been developed to replicate the IBEX ribbon observations and understand the underlying physics \citep[see, e.g.][]{chalov2010, heerikhuisen2010, zirnstein2023}. These models often integrate magnetohydrodynamic (MHD) or kinetic-MHD simulations of the heliosphere with detailed kinetic treatment of ENA production and propagation. Models differ in (a) what populations of primary ENAs are considered and (b) what assumptions are made for pickup proton transport outside the heliopause. 

Currently, there is no consensus on the effectiveness of pitch angle scattering in the outer heliosheath, which depends on the level of turbulence. Some authors assume that pickup protons experience weak pitch angle scattering outside the heliopause, resulting in their highly anisotropic pitch angle distribution \citep[e.g.][]{chalov2010, heerikhuisen2010, moebius2013, zirnstein2013, zirnstein2015b, zirnstein2018}. The other argues that a strong pitch angle scattering may be present either due to PUI self-generated waves or existing Alfvénic turbulence \citep[e.g.][]{florinski2010, schwadron2013, isenberg2014, zirnstein2019}. Several studies have focused on the stability of the pickup ion distribution using linear instability analysis and fully kinetic particle-in-cell and hybrid simulations \citep[e.g.][]{niemiec2016, roytershteyn2019, mousavi2023, mousavi2025}. However, these studies have not yet provided a definitive answer regarding the effectiveness of pitch angle scattering.

In this paper, we investigate the ribbon ENA fluxes produced in the frame of the kinetic-MHD model of the solar wind interaction with the LISM developed by the Moscow group \citep{izmod2015, izmod2020}. In our model, we utilize the secondary charge exchange mechanism of the ribbon production. We consider all relevant primary ENA populations (NSW, NPIs, and ENAs from the IHS) and solve the focused transport equation for pickup proton velocity distribution function in the outer heliosheath in a scatter-free limit (assuming no pitch-angle scattering and no energy diffusion). The results of the simulations are compared with the IBEX-Hi data presented by \citet{mccomas2024}.

The global model simulations of the plasma and H distribution used in this work are described in Section \ref{sec:global_distributions}. Sections \ref{sec:model_atoms} and \ref{sec:model_puis} provide a complete description of the kinetic model used to simulate the distribution of neutralized SW atoms and daughter pickup protons in the outer heliosheath, respectively. Section \ref{sec:model_ribbon} describes the modeling features of the ribbon ENA fluxes observed by the IBEX-Hi instrument. The results of the ribbon simulations and comparison with the IBEX-Hi data are presented in Section \ref{sec:results}. Section \ref{sec:conclusions} provides conclusions and a discussion.

\section{Global distributions of plasma and hydrogen atoms in the heliosphere} \label{sec:global_distributions}

To perform the calculations described in the following sections, the global distributions of plasma and H atoms in the heliosphere must be known. We have carried out global heliospheric simulations of SW/LISM interaction using the kinetic-MHD model described in \citet{izmod2015, izmod2020}, which treats all the charged particles as a single fluid. The main advantage of this model is the use of a flow-adaptive computational grid and exact fitting of discontinuities –- the heliospheric TS in the SW and the HP separating the SW plasma from the interstellar plasma. The global model used in our study enables the simulation of time-dependent distributions, but in this paper, we restrict ourselves to the stationary case.

We utilize the result of the model simulation with the following configuration of the interstellar magnetic field: B$_{\rm LISM}$ = 2.95 $\mu$G, and the angle between $\mathbf{B}_{\rm LISM}$ and $-\mathbf{V}_{\rm LISM}$ equals 40°, where the interstellar magnetic field vector lies in the hydrogen deflection plane \citep[HDP,][]{lallement2005, lallement2010}. \citet{zirnstein2016b} have obtained almost the same configuration by fitting the IBEX ribbon data. The other interstellar parameters of the model are as follows: the hydrogen and proton number densities are n$_{\rm H,LISM}$ = 0.14 cm$^{-3}$ and n$_{\rm p,LISM}$ = 0.04 cm$^{-3}$, the bulk velocity is V$_{\rm LISM}$ = 26.4 km s$^{-1}$, the direction of $\mathbf{V}_{\rm LISM}$ is longitude = 75°.4, latitude = -5°.2 in the ecliptic (J2000) coordinate system, the temperature T$_{\rm LISM}$ = 7500 K was taken, which is slightly higher than 6530 K used in \cite{izmod2020}. At the inner boundary (1 au), the 22-year (1996–-2018) averaged solar cycle conditions were utilized as described in Appendix A of \citet{izmod2020}. The averaged number density and plasma velocity profiles are also shown in panels (B) and (D) of Figure \ref{fig:sw_profile} (see blue solid curves). In the simulation, the charge exchange cross-section from \citet{lindsay2005} was used. This simulation result was also used previously in \citet{florinski2024}.  

Starting from this point, we assume that the distributions of plasma and H atoms are known in the global model simulation domain, which covers both the heliosphere and VLISM. Although this paper presents only the results of stationary simulations, all the equations below are written in a general time-dependent form (with time and its derivatives preserved). Consequently, the model described in this paper can be easily adapted for the time-dependent case. We leave the analysis of time-dependent simulations for future studies.

\section{Kinetic model of the neutralized solar wind atoms distribution} \label{sec:model_atoms}

In this section, we describe the kinetic model used to simulate the distribution of hydrogen atoms originating in the region occupied by the solar wind (primary ENAs). According to the secondary ENA mechanism, these energetic H atoms are grandparents of the IBEX ribbon ENAs.

\subsection{Kinetic equation}

The distribution of hydrogen atoms in the heliosphere is described kinetically. Under the assumption of no elastic collisions, the kinetic equation for the velocity distribution function of hydrogen atoms $f_{\rm H}(t, \mathbf{r}, \mathbf{v})$ can be written as follows:
\begin{align}
    \label{eq:kinetic_atoms}
    & \frac{\partial f_{\rm H}(t, \mathbf{r}, \mathbf{v})}{\partial t} + \mathbf{v} \cdot \frac{\partial f_{\rm H}(t, \mathbf{r}, \mathbf{v})}{\partial \mathbf{r}} + \frac{\mathbf{F}}{m_{\rm H}} \cdot \frac{\partial f_{\rm H}(t, \mathbf{r}, \mathbf{v})}{\partial \mathbf{v}} \\ 
    & = f_{\rm p}(t, \mathbf{r}, \mathbf{v}) \nu_{\rm H}(t, \mathbf{r}, \mathbf{v}) - f_{\rm H}(t, \mathbf{r}, \mathbf{v}) \nu_{\rm ion}(t, \mathbf{r}, \mathbf{v}), \nonumber
\end{align}
where $\mathbf{v}$ is the individual velocity, $m_{\rm H}$ is the mass of H atom, $f_{\rm p}$ is the velocity distribution function of protons, and $\mathbf{F}$ is the force acting on the H atom in the heliosphere, which is the sum of a solar gravitational force $\mathbf{F}_{\rm g}$ and a solar radiative repulsive force $\mathbf{F}_{\rm rad}$. The radiation pressure force is directly proportional to the solar Lyman-$\alpha$ flux, and, therefore, in the general case, $\mathbf{F}_{\rm rad}$ depends on time, the radial velocity of an atom, and heliolatitude. For hydrogen atoms with relatively low velocities (tens of km/s), gravity almost perfectly balances the radiation pressure from the Sun \citep[see, e.g.][]{katushkina2015a, kowalska2020}, so their trajectories can be considered as straight lines. For H atoms with energies of particular interest ($\sim$ few keVs), the influence of radiation pressure on the trajectories is negligible since they are out of the Lyman-$\alpha$ line. However, their kinetic energy is much higher than the potential energy, so the trajectories of atoms are also effectively straight. Therefore, for the sake of simplicity, in our simulations, we assume that all H atoms move in straight lines.

The production rate of H atoms $\nu_{\rm H}$ in equation (\ref{eq:kinetic_atoms}) is
\begin{equation}
    \nu_{\rm H} (t, \mathbf{r}, \mathbf{v}) =  \iiint f_{\rm H}(t, \mathbf{r}, \mathbf{v}_{\rm H}) v_{\rm rel} \sigma_{\rm ex}(v_{\rm rel}) \mathrm{d}\mathbf{v}_{\rm H},
\label{eq:nuH}
\end{equation}
where $v_{\rm rel} = |\mathbf{v} - \mathbf{v}_{\rm H}|$ is the relative proton-atom velocity, and $\sigma_{\rm ex}$ is the charge exchange cross-section. To calculate the production rate of H atoms, we assume that the velocity distribution function of H atoms consists of four components \citep[in the global heliosphere simulations, four different populations depending on the region of their creation, LISM, OHS, IHS, or SSW, are introduced; see, e.g.,][]{izmod2000}. Each velocity distribution function is assumed to be Maxwellian with moments (number density, bulk velocity, and kinetic temperature tensor) taken from the Monte Carlo simulations of the global heliosphere model.

We note that the H atoms velocity distribution function is not Maxwellian \citep{izmod2001}. However, our simulations show that the approximation for $\nu_{\rm H}$, which is the specific moment of the velocity distribution function, using the sum of Maxwellian velocity distributions (with zeroth, first, and second moments taken into account) works reasonably well. This is justified by the facts that according to the global model of the heliosphere (1) the dominant H populations (by their abundance) are so-called primary and secondary interstellar H atoms originating in the LISM and OHS, respectively, and they provide a major contribution to the total production rate, (2) these populations are relatively slow (bulk velocities are $\sim$15--30 km/s) and, more importantly, cold (kinetic temperatures are $\sim$5 000 -- 20 000 K), and (3) the $v_{\rm rel} \sigma_{\rm ex}(v_{\rm rel})$ product is weakly dependent on $v_{\rm rel}$ for energies of particular interest ($\sim$0.1--10 keV). We also note that the Monte Carlo simulations of the global heliosphere provide us not only the moments of the H atom velocity distribution function but also the complete function. However, storing it in the entire simulation domain is not feasible due to memory limitations. 

In equation (\ref{eq:kinetic_atoms}), $\nu_{\rm ion} = \nu_{\rm ex} + \nu_{\rm ph} + \nu_{\rm imp}$ is the total ionization rate due to the ionization processes (charge exchange with protons, photoionization, and electron impact). In our calculations, we neglect electron impact ionization ($\nu_{\rm imp} = 0$), and assume that $\nu_{\rm ph} = \nu_{\rm ph,E} (r_{\rm E} / r)^2$, where $\nu_{\rm ph,E} = 1.67 \times 10^{-7}$s$^{-1}$ is the photoionization rate at $r_{\rm E}$ = 1 au \citep[as it was taken in][]{izmod2015}.
This value of the photoionization rate is meant to represent the solar cycle average. We note that it is slightly overestimated compared to the data presented by e.g. \citet{sokol_etal:20a}. However, we do not expect this to affect the results of our work significantly since, for hydrogen, the photoionization contribution to the total ionization rate is only $\sim$20\%, and the dominant ionization process is charge exchange ($\sim$70\% of the total ionization rate).

The charge exchange ionization rate is
\begin{equation}
    \nu_{\rm ex}(t, \mathbf{r}, \mathbf{v}) = \iiint f_{\rm p}(t, \mathbf{r}, \mathbf{v}_{\rm p}) w_{\rm rel} \sigma_{\rm ex}(w_{\rm rel}) \mathrm{d}\mathbf{v}_{\rm p},
\label{eq:nu_ex1}
\end{equation}
where $w_{\rm rel} = |\mathbf{v} - \mathbf{v}_{\rm p}| = |\mathbf{w} - \mathbf{w}_{\rm p}|$ is the relative atom-proton velocity, $\mathbf{w} = \mathbf{v} - \mathbf{U}$, and $\mathbf{w}_{\rm p} = \mathbf{v}_{\rm p} - \mathbf{U}$ are the velocities in the plasma reference frame that is moving with the bulk velocity $\mathbf{U}$. We assume that the proton velocity distribution function is isotropic in the plasma reference frame, so $\hat f_{\rm p}(t, \mathbf{r}, w_{\rm p}) \equiv f_{\rm p}(t, \mathbf{r}, \mathbf{U} + \mathbf{w}_{\rm p})$. It can be shown that, in this case, the charge exchange ionization rate depends only on the module of $\mathbf{w}$:
\begin{equation}
    \nu_{\rm ex}(t, \mathbf{r}, w) = \int_0^{\infty} \hat f_{\rm p}(t, \mathbf{r}, w_{\rm p}) I(w, w_{\rm p}) w_{\rm p}^2 \mathrm{d}w_{\rm p},
\label{eq:nu_ex2}
\end{equation}
where
\begin{align}
    I(w, w_{\rm p}) 
    & = \iint w_{\rm rel} \sigma_{\rm ex}(w_{\rm rel}) \sin \vartheta \mathrm{d}\vartheta \mathrm{d}\phi \label{eq:charge_exchange_integral} \\
    & = 2\pi\int_0^{\pi} u_{\rm rel} \sigma_{\rm ex}(u_{\rm rel}) \sin \vartheta \mathrm{d}\vartheta, \nonumber
\end{align}
($w_{\rm p}$, $\vartheta$, $\phi$) are components of $\mathbf{w}_{\rm p}$ in the spherical coordinate system, and $u_{\rm rel} = \sqrt{w^2 - 2 w w_{\rm p} \cos \vartheta + w_{\rm p}^2}$. Before our simulations, we calculated the values of integral (\ref{eq:charge_exchange_integral}) for different $w$ and $w_{\rm p}$ and stored them in the lookup table.

\subsection{Method of characteristics}

The kinetic equation (\ref{eq:kinetic_atoms}) is a linear partial differential equation, and it can be solved by a method of characteristics. A characteristic is a curve in the phase space $(\mathbf{r}, \mathbf{v})$. It is defined by the system of equations:
\begin{equation}
    \frac{\mathrm{d} \mathbf{r}}{{\rm d} t} = \mathbf{v},
    \frac{\mathrm{d} \mathbf{v}}{\mathrm{d} t} = \frac{\mathbf{F}}{m_{\rm H}},
    \label{eq:characteristic_atom}
\end{equation}
which describes the trajectory of an H atom. Along the characteristic, the kinetic equation (\ref{eq:kinetic_atoms}) transforms to
\begin{equation}
    \frac{\mathrm{d} f_{\rm H}}{{\rm d} t} = f_{\rm p} \nu_{\rm H} - f_{\rm H} \nu_{\rm ion},
    \label{eq:kinetic_atom_short}
\end{equation}
and the solution can be written as:
\begin{align}
    \label{eq:solution_atom}
    f_{\rm H}(t, \mathbf{r}, \mathbf{v}) 
    & = f_{\rm H,0}(t_0, \mathbf{r}_0, \mathbf{v}_0) L_{\rm H}(t_0, t) \\ 
    & + \int_{t_0}^{t} f_{\rm p}(\tau, \mathbf{r}(\tau), \mathbf{v}(\tau)) \nu_{\rm H}(\tau, \mathbf{r}(\tau), \mathbf{v}(\tau)) L_{\rm H}(\tau, t) {\rm d} \tau, \nonumber 
\end{align}
where $f_{\rm H,0}$ is the velocity distribution function of H atoms at a far away from the Sun at the moment $t_0$ in the past, and $L_{\rm H}$ is the factor that describes the loss of H atoms due to ionization processes:
\begin{equation}
    L_{\rm H}(\tau, t) = \exp\left(- \int^t_{\tau} \nu_{\rm ion}(\tau^*, \mathbf{r}(\tau^*), \mathbf{v}(\tau^*)) {\rm d} \tau^* \right).
\end{equation}
The integration in equation (\ref{eq:solution_atom}) is performed backward in time along the atom trajectory from the point $(t, \mathbf{r}, \mathbf{v})$ to the point $(t_0, \mathbf{r}_0(t_0), \mathbf{v}_0(t_0))$, which are connected by the system of equations (\ref{eq:characteristic_atom}). In this work, we are particularly interested in the H atoms of solar wind origin, so we adopt $f_{\rm H,0} \equiv 0$ at the outer computational boundary (1000 au from the Sun in our simulations).

\subsection{Hydrogen populations of solar wind origin}\label{sec:hpopulations}

We consider five populations of H atoms dependent on the region of creation and their parent proton population, which have different properties, and simulate their distributions separately (see Table \ref{tab:Hpopulations} for description of populations). The H atoms of Types 1.0 and 1.1 are born in the supersonic solar wind region (Region 1), while atoms of Types 2.0, 2.1, and 2.2 originate in the inner heliosheath (Region 2). Therefore, for atoms of Types 1.0 and 1.1, the velocity distribution function of parent protons in equation (\ref{eq:solution_atom}) is assumed to be non-zero only inside the termination shock, and for atoms of Types 2.0, 2.1, and 2.2 -- only in the inner heliosheath. Note that, while reconstructing the trajectory of primary ENA of Type 2.0, 2.1, or 2.2 backward-in-time from a point in the outer heliosheath, there may be one or two trajectory sections within the inner heliosheath. The case of two sections corresponds to the situation when the trajectory intersects the SSW region.

\begin{table*}
    \centering
    \caption{Different hydrogen populations of the solar wind origin (dependent on the region of their creation and parent proton population).}
    \label{tab:Hpopulations}
    \begin{tabular}{llll} 
        \hline
        Type & Region of creation & Parent proton population & Abbreviation\\
        \hline
        1.0 & Supersonic solar wind (1) & Thermal solar wind protons & Neutral Solar Wind (NSW)\\
        1.1 & Supersonic solar wind (1) & Pickup protons originated in Region 1 & Neutralized Pickup Ions (NPIs)\\
        2.0 & Inner heliosheath (2) & Thermal solar wind protons & ENAs from SW protons\\
        2.1 & Inner heliosheath (2) & Pickup protons originated in Region 1 & ENAs from SSW PUIs\\
        2.2 & Inner heliosheath (2) & Pickup protons originated in Region 2 & ENAs from IHS PUIs\\
        \hline
    \end{tabular}
\end{table*}

Atoms of Types 1.0 and 2.0 originate in the charge exchange process of thermal solar wind protons, while Types 1.1 and 2.1 are the result of charge exchange of pickup protons that are born in the supersonic solar wind. We also consider separately the population of atoms that originate from the pickup protons that are born in the inner heliosheath (Type 2.2). Depending on the parent proton population -- thermal solar wind protons or pickup protons -- we use different velocity distribution functions of protons in equation (\ref{eq:solution_atom}) -- $f_{\rm p} = f_{\rm sw}$ or $f_{\rm p} = f_{\rm pui}$, respectively. To calculate the velocity distribution functions of pickup and thermal solar wind protons, we use the kinetic model developed by \citet{baliukin2020, baliukin2023}, which enables accounting for thermal and suprathermal proton populations separately. To simulate the energetic population of pickup protons that are accelerated at the heliospheric termination shock (low-energy part of the anomalous cosmic ray spectrum), we use the power-law-tail scenario \citep[see][]{baliukin2022}.

We also note that our model does not capture the non-adiabatic behavior of thermal proton temperature in the supersonic solar wind region, which is observed by Voyager 2 \citep[see, e.g.][]{gazis1994,lazarus1995}. The dissipation of waves generated by the unstable ring distribution of pickup ions and ambient turbulence has been proposed as the main physical process responsible for the solar wind temperature profile \citep{williams1995, isenberg2003}. However, recently, \citet{korolkov2022} showed that heating of the solar wind due to the passage of the shocks or shock layers is the dominant mechanism that leads to the non-adiabatic behavior of the solar wind temperature. They used a time-dependent data-driven model that employs the solar-wind parameters at 1 AU with minute resolution as the boundary condition (based on the OMNI data). To capture the non-adiabatic behavior of thermal proton temperature, in our stationary simulations we assume the temperature to be equal to 10 000 K if the model provides a lower value. 

Atoms of Types 2.0, 2.1, and 2.2 (a) originate in Region 2, where the solar wind bulk velocity is low (down to zero at the stagnation point) and (b) have relatively high thermal velocities. Therefore, these atoms can propagate both inward to the vicinity of the Sun and outward outside the heliopause from the inner heliosheath. The fluxes of those that move closer to the Sun are usually called Globally Distributed Fluxes (GDF), and they were the subject of many previous studies \citep[see, e.g., reviews by][]{galli2022, sokol2022}. In this work, we are particularly interested in those atoms that leave the heliosphere since they are the seed population for pickup protons in the outer heliosheath.

\begin{figure*}
\includegraphics[width=\textwidth]{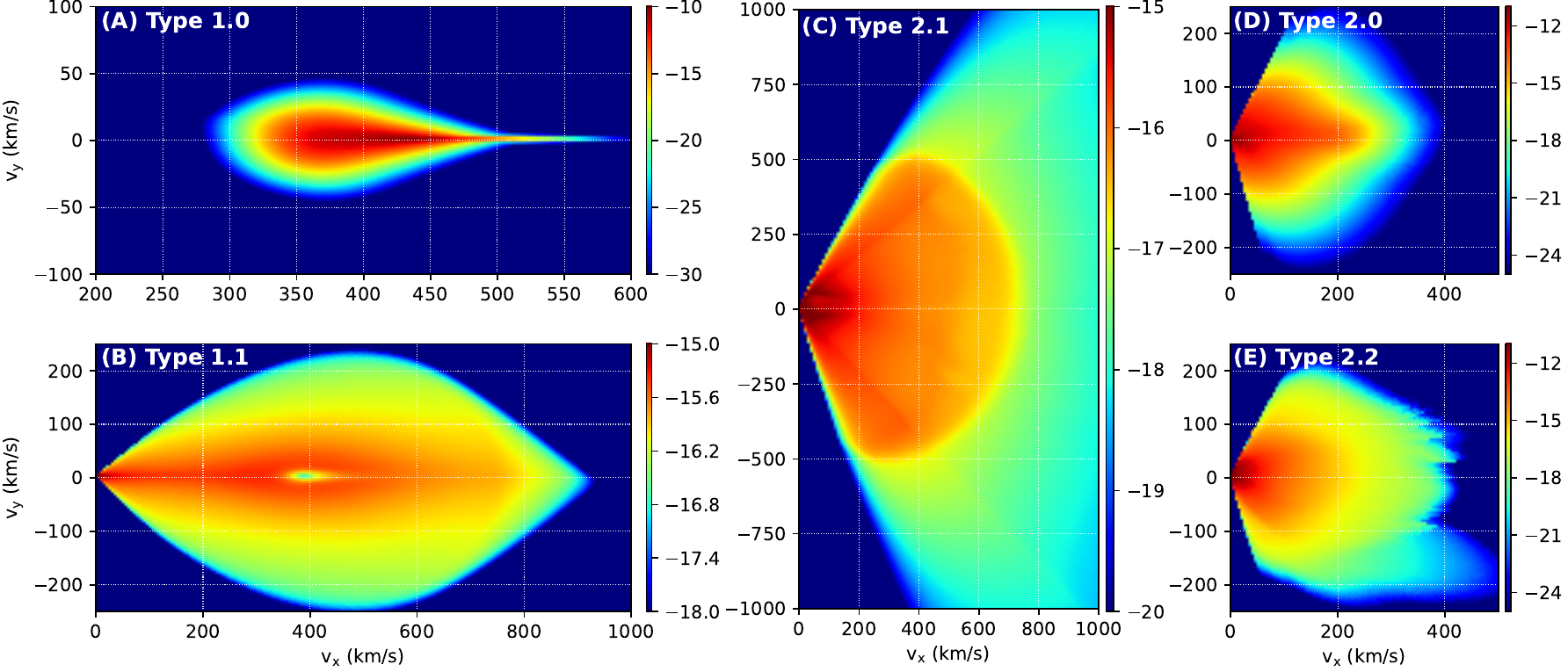}
\caption{
Contour plots of the velocity distribution function's $f_{\rm H}$ (in s$^3$/m$^6$) decimal logarithm of different Types of hydrogen atoms (described in Table \ref{tab:Hpopulations}) at 150 au in the upwind direction. The slices are shown as a function of the $v_{\rm x}$ and $v_{\rm y}$ components (in km/s) for $v_z = 0$.
}
\label{fig:Hpopulations_2D}
\end{figure*}

\begin{figure}
\includegraphics[width=\columnwidth]{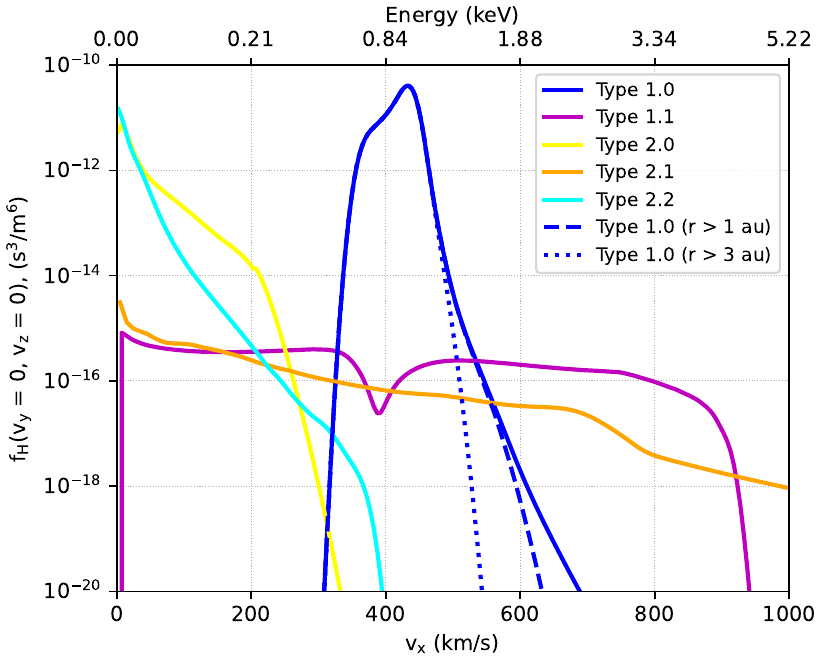}
\caption{
One-dimensional slices of the velocity distribution function (at 150 au in the upwind direction) of different hydrogen components as functions of $v_x$ (in km/s) for $v_y = 0$ and $v_z = 0$. The blue, magenta, yellow, orange, and cyan solid lines show Types 1.0, 1.1, 2.0, 2.1, and 2.2 velocity distribution profiles, respectively. The blue dashed and dotted lines show Type 1.0 velocity distribution functions with the contribution of sources outside 1 au and 3 au from the Sun, respectively.
}
\label{fig:Hpopulations_1D}
\end{figure}

Figures \ref{fig:Hpopulations_2D} and \ref{fig:Hpopulations_1D} show the velocity distribution functions of different populations of atoms described above at 150 au in the upwind direction (in the outer heliosheath). The system of coordinates is aligned with the interstellar flow and magnetic field vector: the $X$-axis is directed toward the interstellar flow, i.e., opposite to $\mathbf{V}_{\rm LISM}$ (upwind direction), the $Y$-axis is in the plane that contains $\mathbf{V}_{\rm LISM}$ and $\mathbf{B}_{\rm LISM}$ vectors (BV--plane) and is perpendicular to the X-axis such that the $Y$-axis has a positive projection on the north solar pole direction, and the $Z$-axis complements the right-handed triplet. In Figure \ref{fig:Hpopulations_2D}, the velocity distributions are presented for $v_z = 0$ as functions of the $v_x$ and $v_y$ components (2D slices), while Figure \ref{fig:Hpopulations_1D} provides the one-dimensional slices as functions of $v_x$ (for $v_y = 0$ and $v_z = 0$). 

As can be seen from Figure \ref{fig:Hpopulations_2D}(A) and the blue solid line in Figure \ref{fig:Hpopulations_1D}, the Neutral Solar Wind (Type 1.0) has a very narrow, radially oriented velocity distribution function. These atoms have a bulk velocity of $\sim$430 km/s, corresponding to the slow solar wind (the upwind direction is only a few degrees from the solar equatorial plane). There is also a notable asymmetry in the blue solid line of Figure \ref{fig:Hpopulations_1D}, induced by the fact that NSW atoms originate at different distances from the Sun, and the solar wind is slowing down with increasing distance (due to charge exchange with interstellar atoms). 

One more distinctive feature of the NSW velocity distribution seen in Figure \ref{fig:Hpopulations_2D}(A) is the beam-like enhancement at $v_y = 0$ (for velocities higher and lower than the solar wind bulk velocity). The atoms that comprise this beam originate close to the Sun. As seen in Figure \ref{fig:Hpopulations_1D}, the main contribution to the NSW comes from distances greater than 3 au (compare blue solid and dotted lines). To simulate sources within 1 au from the Sun, which is the inner boundary in the global heliosphere model, we extrapolate the solar wind proton parameters assuming (1) the number density $n_{\rm sw} \propto 1 / r^2$, (2) the temperature $T_{\rm sw} \propto 1 / r^{2\gamma-2}$ (adiabatic law, $\gamma$ = 5/3), and (3) a linear dependence of the velocity on the radial distance. The comparison of the solid and dashed blue lines in Figure \ref{fig:Hpopulations_1D} shows that the contribution to the NSW of sources within 1 au from the Sun can be considered negligible.

The velocity distribution of NPIs (Type 1.1) shown in Figures \ref{fig:Hpopulations_2D}(B) and \ref{fig:Hpopulations_1D} (magenta curve) is very different from the NSW velocity distribution. While generally, it preserves the filled-shell-like distribution of the parent pickup protons, the distribution function is compiled from the different PUI shells that contribute through charge exchange along the trajectory. As can be seen from the magenta curve of Figure \ref{fig:Hpopulations_1D}, some NPIs have much higher velocities than NSW atoms (up to twice the solar wind speed). Another feature is the dip at $\sim$400 km/s -- a heritage of the filled-shell distribution function that approaches zero at low velocities in the solar wind reference frame. The velocity distributions of the NSW and NPI components are studied based on the kinetic model and discussed in detail in \citet{florinski2017}. The properties of atoms that originate in the supersonic solar wind region (a mixture of NSW and NPIs) were also studied by \citet{heerikhuisen2016} and \citet{gamayunov2017}. 

We note that the acute angle on the left side of Figure \ref{fig:Hpopulations_2D}(B) defines the range of velocities and corresponding trajectories that, while being traced backward from 150 au, cross the heliospheric termination shock (the trajectories with velocities within the blue region outside this angle do not cross the termination shock). The angle on the left side of panels (C-E) in Figure \ref{fig:Hpopulations_2D} (Types 2.0, 2.1, and 2.2 that originate in the inner heliosheath) is wider than in panel (B) since the heliopause covers a wider solid angle than the heliospheric termination shock (as seen from 150 au upwind). The slight asymmetry in $v_y$ seen in panels (B) and (C-E) is due to the asymmetry of the heliospheric termination shock and heliopause, respectively. It is caused by the inclined interstellar magnetic field that pushes the heliosphere with the interstellar magnetic pressure maximum from the southwest (negative $Y$ coordinates in the BV-plane), where the magnetic field lines are parallel to the heliopause surface.

Figure \ref{fig:Hpopulations_2D}(C) and the orange curve in Figure \ref{fig:Hpopulations_1D} show the velocity distribution function of Type 2.1. As can be seen, this component has a broad distribution, inherited from the parent pickup protons in the inner heliosheath. The distinctive feature of this distribution is the high-velocity tail seen at velocities higher than $\sim$800 km/s. This tail is produced by the acceleration of the parent pickup protons at the termination shock. This population of protons is described by a power-law-tail scenario for the downstream velocity distribution function used in this work \citep{baliukin2022}.

Types 2.0 and 2.2 look rather similar with relatively low energy and cold velocity distributions with respect to other populations, as can be seen in panels (D-E) in Figure \ref{fig:Hpopulations_2D} and yellow and cyan curves in Figure \ref{fig:Hpopulations_1D}. Type 2.0 is the product of charge exchange of solar wind thermal protons, which have the temperature (thermal velocity) of $\sim$50 000 K ($\sim$30 km/s) of the inner heliosheath \citep[according to Voyager 2 observations, e.g.][]{richardson2022}, and whose distribution is assumed to be Maxwellian. The population of pickup protons originating in the inner heliosheath has a low bulk velocity and low thermal velocity (and so do their daughter atoms of Type 2.2), since the relative proton -- interstellar atom velocity at the moment of pickup is low in this region.

\section{Kinetic model of pickup proton distribution outside the heliopause} \label{sec:model_puis}

In the outer heliosheath beyond the HP, the energetic atoms of solar wind origin experience charge exchange and form a suprathermal population of pickup protons. According to the secondary ENA mechanism discussed in Section \ref{sec:intro}, those pickup protons in the outer heliosheath are the parents of the IBEX ribbon ENAs. In this section, we describe the kinetic model used to simulate the distribution of pickup protons in the outer heliosheath. 

\subsection{Anisotropic transport of pickup protons}\label{sec:anisotropic_transport}

The gyroperiod and gyroradius of pickup protons are generally much smaller than other time and spatial scales ($T_{\rm g} \lesssim 10^{4}$ s, $r_{\rm g} \lesssim 10^{-3}$ au). Thus, their velocity distribution function can be considered gyrotropic \citep{chalov2006}. The gyrophase-averaged velocity distribution function of pickup protons in the plasma reference frame (moving with velocity $\mathbf{U}$) is
\begin{equation}
    f_{\mathrm{pui}}^*(t, \mathbf{r}, w, \mu) = \frac{1}{2 \pi} \int_0^{2 \pi} f_{\mathrm{pui}}(t, \mathbf{r}, \mathbf{U} + \mathbf{w}) \mathrm{d} \varphi.
\end{equation}
The components of a pickup proton's velocity in the plasma rest frame $\mathbf{w} = \mathbf{v} - \mathbf{U}$ along and perpendicular to the magnetic field line are 
\begin{equation}
w_{\parallel} = w \mu,\: w_{\perp1} = w \sqrt{1 - \mu^2} \cos\varphi,\: w_{\perp2} = w \sqrt{1 - \mu^2} \sin\varphi,
\end{equation}
where $\mu = \cos\xi = \mathbf{w} \cdot \mathbf{b} / {w}$ is the cosine of the angle $\xi$ between the particle velocity in the plasma rest frame and magnetic field (the so-called pitch-angle), $\mathbf{b} = \mathbf{B} / B$ is the unit vector of the magnetic field, and $\varphi$ is the gyrophase.

The gyrophase-averaged velocity distribution function obeys the Fokker-Planck type focused transport equation \citep[see, e.g.][]{skilling1971, isenberg1997}:
\begin{equation}
    \frac{\partial f_{\rm pui}^*}{\partial t} + \mathbf{V} \cdot \frac{\partial f_{\rm pui}^*}{\partial \mathbf{r}} + A_w \frac{\partial f_{\rm pui}^*}{\partial w} + A_{\mu} \frac{\partial f_{\rm pui}^*}{\partial \mu} = \hat D f_{\rm pui}^* + S,
    \label{eq:kinetic_anisotropic}
\end{equation}
The following notation was introduced:
\begin{align}
    & \mathbf{V} = \mathbf{U} + w \mu \mathbf{b}, \\
    & A_w = w \left( 
    \frac{1 - 3 \mu^2}{2} \mathbf{b}^{\rm T} \frac{\partial \mathbf{U}}{\partial \mathbf{r}} \mathbf{b} 
    - \frac{1 - \mu^2}{2} \mathrm{div}\mathbf{U}
    - \frac{\mu \mathbf{b}}{w} \cdot \frac{\mathrm{d} \mathbf{U}}{\mathrm{d} t}
    \right), \label{eq:Aw}\\
    & A_{\mu} = \frac{1 - \mu^2}{2}  \left(
    w \mathrm{div}\mathbf{b} + \mu \mathrm{div}\mathbf{U}
    - 3 \mu \mathbf{b}^{\rm T} \frac{\partial \mathbf{U}}{\partial \mathbf{r}} \mathbf{b}
    - \frac{\mathbf{b}}{w} \cdot \frac{\mathrm{d} \mathbf{U}}{\mathrm{d} t}
    \right), \label{eq:Amu}
\end{align}
The second term on the left side of the equation (\ref{eq:kinetic_anisotropic}) describes convective transport with the solar wind velocity and the motion of protons along magnetic field lines due to the anisotropy of their pitch-angle distribution, while the third and fourth terms describe the processes of adiabatic energy change and focusing in a non-uniform magnetic field, respectively. 

The source/sink term $S$ on the right side of the equation (\ref{eq:kinetic_anisotropic}) has the following general form: $S = S_{+} - S_{-} f_{\rm pui}^*$, where functions $S_{+}$ and $S_{-}$ are responsible for production and losses (extinction) of pickup protons due to charge exchange processes, respectively:
\begin{align}
    \label{eq:s1term}
    S_+(t, \mathbf{r}, w, \mu) 
    & = \frac{1}{2 \pi} \int_0^{2 \pi} f_{\rm H}(t, \mathbf{r}, \mathbf{U} + \mathbf{w}) \nu_{\rm ion}(t, \mathbf{r}, \mathbf{U} + \mathbf{w}) \mathrm{d} \varphi \\
    & = \frac{\nu_{\rm ex}(t, \mathbf{r}, w) + \nu_{\rm ph}(r)}{2 \pi} \int_0^{2 \pi} f_{\rm H}(t, \mathbf{r}, \mathbf{U} + \mathbf{w}) \mathrm{d} \varphi \nonumber \\
    S_-(t, \mathbf{r}, w, \mu) &= \frac{1}{2 \pi} \int_0^{2 \pi} \nu_{\rm H}(t, \mathbf{r}, \mathbf{U} + \mathbf{w}) \mathrm{d} \varphi.
\label{eq:s2term}
\end{align}
In connection with the five populations of atoms introduced in Section \ref{sec:hpopulations}, we consider different populations of daughter pickup protons in the outer heliosheath in our simulations. To be more specific, to calculate the velocity distribution function of pickup protons that originate from a particular population of solar wind H atoms, in the source term (\ref{eq:s1term}) we use the corresponding velocity distribution function $f_{\rm H}$ of these H atoms defined by equation (\ref{eq:solution_atom}).

On the right side of equation (\ref{eq:kinetic_anisotropic}), $\hat D$ is the scattering operator that describes the pitch-angle scattering and energy diffusion. The general form of the operator is \citep[e.g.][]{schlickeiser1989}
\begin{align}
    \hat D f_{\rm pui}^* 
    &= \frac{\partial}{\partial \mu} \left( D_{\mu \mu} \frac{\partial f_{\rm pui}^*}{\partial \mu} \right) + \frac{\partial}{\partial \mu} \left( D_{\mu w} \frac{\partial f_{\rm pui}^*}{\partial w} \right) \\
    & + \frac{1}{w^2} \frac{\partial}{\partial w} \left( w^2 D_{\mu w} \frac{\partial f_{\rm pui}^*}{\partial \mu} \right) 
    + \frac{1}{w^2} \frac{\partial}{\partial w} \left( w^2 D_{w w} \frac{\partial f_{\rm pui}^*}{\partial w} \right), \nonumber
\end{align}
where $D_{\mu \mu}$, $D_{\mu w}$, and $D_{w w}$ are the diffusion coefficients.

In this work, we adopt a simple model with no pitch-angle scattering and no energy diffusion, i.e., $\hat D f_{\rm pui}^* \equiv 0$. This assumption implies that there is (a) a low level of preexisting turbulence outside the heliopause and (b) no PUI-generated instabilities. It is important to note that the narrow PUI ring distribution rapidly excites instabilities, leading to near-isotropic distribution within days, which is much shorter than the charge exchange timescale $1 / \nu_{\rm H}$ ($\sim$few years for energies under consideration), as first pointed out by \citet{florinski2010} based on hybrid simulations. Such fast isotropization would suppress the ribbon’s narrow structure, raising doubt on the viability of the weak-scattering scenario. However, a recent study by \citet{mousavi2025} using realistic multi-component PUI distributions shows that near $90^\circ$ pitch angles -- the ribbon direction -- the scattering timescale by the Alfvén-cyclotron waves is several years (based on the estimations of pitch angle diffusion coefficient). This outcome is due to the mirror waves that reduce the thermal anisotropy, indirectly suppressing instability growth. Their result aligns with the weak-scattering scenario. We plan to ease the scatter-free assumption made in our simulations and adopt more realistic diffusion coefficients in future work.

\subsection{Method of characteristics}

In case of no diffusion, the kinetic equation (\ref{eq:kinetic_anisotropic}) is a linear first-order partial differential equation, so the method of characteristics can be used to solve it for the velocity distribution function of pickup protons (analogous to the previous section). A characteristic in this case is a curve in the phase space $(\mathbf{r}, w, \mu)$ of coordinates, velocity, and pitch-angle cosine, which is defined by the system of equations:
\begin{equation}
    \frac{\mathrm{d} \mathbf{r}}{{\rm d} t} = \mathbf{V},
    \frac{\mathrm{d} w}{\mathrm{d} t} = A_w,
    \frac{\mathrm{d} \mathbf{\mu}}{{\rm d} t} = A_{\mu}.
    \label{eq:characteristic}
\end{equation}

Along the characteristic (\ref{eq:characteristic}), the transport equation (\ref{eq:kinetic_anisotropic}) transforms to the following ordinary differential equation:
\begin{equation}
    \frac{\mathrm{d} f_{\rm pui}^*}{{\rm d} t} = S_{+} - S_{-} f_{\rm pui}^*.
    \label{eq:kinetic_short}
\end{equation}
Assuming that $f_{\rm pui}^* \equiv 0$ at far distances from the Sun (in the LISM), the solution of equation (\ref{eq:kinetic_short}) can be written as
\begin{equation}
    f_{\rm pui}^{*}(t, \mathbf{r}, w, \mu) = \int^t_{t_0} S_+(\tau, \mathbf{r}(\tau), w(\tau), \mu(\tau)) L_{\rm pui}(\tau, t) {\rm d} \tau,
    \label{eq:solution}
\end{equation}
where $L_{\rm pui}$ is the factor that describes the loss of pickup protons due to the charge exchange with hydrogen atoms:
\begin{equation}
    L_{\rm pui}(\tau, t) = \exp\left(- \int^t_{\tau} S_-(\tau^*, \mathbf{r}(\tau^*), w(\tau^*), \mu(\tau^*)) {\rm d} \tau^* \right).
\end{equation}

Therefore, to calculate the velocity distribution function, we reconstruct the phase-space characteristic curve (\ref{eq:characteristic}) backward in time and integrate equation (\ref{eq:solution}) until we reach at some moment $\tau = t_0$ the outer computational boundary, which is 1000 au from the Sun in our simulations, where $f_{\rm pui}^* \equiv 0$ is assumed.

\subsection{Analysis of pickup proton dynamics}

For simplicity, let us consider a case of no drift motion associated with plasma flow outside the heliopause, i.e., $\mathbf{U} = 0$. Therefore, the characteristics of the kinetic equation (\ref{eq:kinetic_anisotropic}) in the physical space coincide with the magnetic field lines with $\mathrm{d}s / \mathrm{d}t = V = w \mu$, where $s$ is the coordinate along the magnetic field line. We note that this assumption is well justified almost everywhere since $U \approx$ 20 km/s in the OHS, and $U \ll w$ for IBEX-Hi energies. However, in the regions where $\mu \approx 0$, which are of particular interest, the drift velocity should be taken into account.

Along the magnetic field line, we have $\mathrm{d} w / \mathrm{d} t = A_{\rm w} = 0$, i.e., $w = {\rm const}$, and
\begin{equation}
    \frac{\mathrm{d} \mu}{\mathrm{d} t} = \frac{1 - \mu^2}{2} w \mathrm{div}\mathbf{b}.
    \label{eq:dmu_dt_divb}
\end{equation}
Integration of the last equation along the particular magnetic field line and consideration of the formula 
\begin{equation}
    \mathrm{div}\mathbf{b} = -\frac{1}{B} \frac{\partial B}{\partial s},
    \label{eq:divb}
\end{equation}
which follows from $\mathrm{div}\mathbf{B} = 0$, provides the conservation law: $(1 - \mu^2) / B = {\rm const}$. 

Accounting for $w = {\rm const}$, as a result, we have the following conservation laws along the magnetic field line:
\begin{equation}
    \frac{w_{\perp}^2}{B} = {\rm const}, \: w_{\perp}^2 + w_{\parallel}^2 = {\rm const},
    \label{eq:conservation_laws}
\end{equation}
where $w_{\parallel} = w \mu$ and $w_{\perp} = w \sqrt{1 - \mu^2}$. The first equation here is the magnetic moment (first adiabatic invariant), while the second equation states the conservation of particle energy. One can see that if a particle moves along the field line in the direction of increasing magnetic field, its parallel velocity becomes smaller, and vice versa. The parallel velocity of some particles may even become zero, so their direction of motion changes (magnetic mirror effect). So, the pickup protons spend a comparatively long time in the regions near the magnetic field maxima, which makes these regions favorable for secondary ENA production. 

\citet{chalov2010} used the system of equations (\ref{eq:conservation_laws}) to capture the dynamics of protons beyond the heliopause. They utilized the simplified guiding center approach, neglecting the drift velocity $\mathbf{U}$. In our simulations, we do not use such simplification. The other significant improvement of our model is that in addition to NSW (Type 1.0), we take into account also the other components of primary ENAs, namely NPIs (Type 1.1) and ENAs from the IHS (Types 2.0, 2.1, and 2.2), which we consider kinetically.

We note that the simple qualitative analysis of pickup proton dynamics performed in this section assumes a scatter-free limit. However, several previous studies challenged this assumption \citep[e.g.][]{florinski2010, mousavi2023}, as we mentioned in Section \ref{sec:anisotropic_transport}.

\section{Modeling of the IBEX ribbon fluxes} \label{sec:model_ribbon}

To simulate the velocity distribution function of the IBEX ribbon ENAs produced by the secondary ENA mechanism, the equation (\ref{eq:solution_atom}) can be utilized again. Secondary ENAs are daughters to pickup protons outside the heliopause, so some modifications to this equation are made: (a) the velocity distribution function of parent protons $f_{\rm p} = f_{\rm pui}^*$, which comes from the solution of equation (\ref{eq:kinetic_anisotropic}) is utilized (it is non-zero only outside the heliopause), and (b) $f_{\rm H,0}(t_0) = 0$ is used at far distances from the Sun, where we assume that there are no atoms of the ribbon origin (the outer boundary in the calculations is 1000 au). Therefore, the solution for the velocity distribution function $f_{\rm H, RIB}$ of the IBEX ribbon ENAs has the following form:
\begin{equation}
    f_{\rm H,RIB}(t, \mathbf{r}, \mathbf{v}) = \int_{t_0}^t f_{\rm pui}^{*} \nu_{\rm H}(\tau, \mathbf{r}(\tau), \mathbf{v}(\tau)) L_{\rm H}(\tau, t) {\rm d} \tau,
    \label{eq:fHRIB}
\end{equation}
where $f_{\rm pui}^* = f_{\rm pui}^{*}(\tau, \mathbf{r}(\tau), w(\tau), \mu(\tau))$ is the velocity distribution function of pickup protons in the outer heliosheath, $\mathbf{w}(\tau) = \mathbf{v}(\tau) - \mathbf{U}(\tau)$ and $\mu(\tau) = \mathbf{w}(\tau) \cdot \mathbf{b}(\tau) / w(\tau)$ are local parent pickup proton velocity and the cosine of the pitch-angle in the plasma reference frame at the point of H atom trajectory $\mathbf{r} = \mathbf{r}(\tau)$.

The differential flux $j$ of ribbon H atoms in the direction of a line of sight (LOS) as seen by the moving observer is
\begin{equation}
    j(t_{\rm obs}, \mathbf{r}_{\rm obs}, E_{\rm rel}, \mathbf{e}_{\rm los}) = \frac{2}{m_{\rm H}^2} f_{\rm H, RIB}(t_{\rm obs}, \mathbf{r}_{\rm obs}, \mathbf{v}) E_{\rm rel},
\label{eq:flux}
\end{equation}
$t_{\rm obs}$ is the moment of observation, $\mathbf{r}_{\rm obs}$ is the position of the observer (IBEX spacecraft at 1 au from the Sun), $\mathbf{v} = \mathbf{v}_{\rm obs} + \mathbf{v}_{\rm rel}$ is the velocity of H atom in the Sun reference frame, $\mathbf{v}_{\rm obs}$ is the velocity of the observer in the Sun reference frame, $\mathbf{v}_{\rm rel} = -v_{\rm rel} \mathbf{e}_{\rm los}$ is the velocity of the H atom relative to the observer, $\mathbf{e}_{\rm los}$ is the unit vector in the direction of line of sight, and $E_{\rm rel} = m_{\rm H} v_{\rm rel}^2 / 2$ is the relative energy registered by the instrument (IBEX-Hi).

In this study, we use the data of IBEX ram observations with the survival probability and Compton-Getting corrections applied (see Section \ref{sec:comparison}). So, to make our simulation results directly comparable with IBEX data, we (a) consider the lines of sight when the sensor views the heliosphere in the ram direction (i.e. when the angle between the LOS and the direction of the spacecraft's motion is less than 90$^\circ$), (b) ignore the ionization losses of ribbon H atoms within 100 au from the Sun (in $L_{\rm H}$ term of equation \ref{eq:fHRIB}), and (c) do not take into account the motion of the IBEX spacecraft (assume $v_{\rm obs} = 0$ in equation \ref{eq:flux}).

To calculate the differential flux $J_{\rm i}$ in the LOS direction measured by {\it IBEX-Hi} instrument at the energy channel \#i (where i = 2, ..., 6, and center energies are $\sim$0.71, 1.11, 1.74, 2.73, and 4.29 keV, respectively), the energy transmission of the electrostatic analyzer (ESA) is taken into account:
\begin{equation}
J_{\rm i}(t_{\rm obs}, \mathbf{r}_{\rm obs}, \mathbf{e}_{\rm los}) = \int^{E_{\rm i+}}_{E_{\rm i-}}  j(t_{\rm obs}, \mathbf{r}_{\rm obs}, E_{\rm rel}, \mathbf{e}_{\rm los}) T_{\rm i}(E_{\rm rel}) {\rm d}E_{\rm rel},
\label{eq:flux_energy_resp} 
\end{equation}
where $E_{\rm i-}$ and $E_{\rm i+}$ define the range of ESA \#i accepting energies, $T_{\rm i}(E_{\rm rel})$ is the normalized energy response function of the ESA \#i \citep[see, e.g., Appendix A in][]{baliukin2020} such as $\int_{E_{\rm i-}}^{E_{\rm i+}} T_{\rm i}(E_{\rm rel}) {\rm d}E_{\rm rel} = 1$. Accounting for energy transmission leads to a redistribution of the fluxes between the energy channels and a smoothing of the observed energy spectrum.

Since the ribbon is a relatively narrow structure in the sky, it is necessary to account for the angular smoothing effects of the {\it IBEX-Hi} collimator in the modeling, specifically if a ribbon width analysis is performed. To do so in our simulations, we consider the point spread function of the {\it IBEX-Hi} sensor \citep{funsten2009} and approximate its hexagonal shape with a circular shape, i.e., we assume that the detection probability depends only on the angle between the incident H atom velocity and collimator axis. To produce the full-sky maps, firstly, we simulate the model maps with $2^\circ \times 2^\circ$ pixel resolution (the fluxes are calculated for pixel centers). For the corresponding IBEX $6^\circ \times 6^\circ$ pixels, the resulting flux is calculated as a weighted average of the nearby pixel fluxes using the collimator point spread function.

Here, we briefly summarize the method used to calculate the secondary ENA flux observed by IBEX-Hi at a specific time, position, and energy. In our modeling, we utilize the method of characteristics, as described in previous sections, with backward-in-time integration. We start our calculations from the observer position. The simulation process consists of three steps.
\begin{enumerate}
    \item Secondary ENA integration step. To calculate the value of the velocity distribution function of secondary ENAs $f_{\rm H, RIB}$ in equation (\ref{eq:flux}), we integrate the production rate along the secondary ENA characteristic, which coincides with the LOS in the physical space (equation \ref{eq:fHRIB}).
    \item PUI integration step. For each point along the secondary ENA characteristic in the outer heliosheath, we calculate the velocity distribution function of PUIs $f_{\rm pui}^{*}$ (using equation \ref{eq:solution}), which itself is the integral along the PUI characteristic defined by a set of equations (\ref{eq:characteristic}).
    \item Primary ENA integration step. The PUI production rate $S_+$ is the integral of the velocity distribution function of primary ENAs $f_{\rm H}$. To calculate $f_{\rm H}$, we integrate the equation (\ref{eq:solution_atom}) along the primary ENA characteristic defined by a set of equations (\ref{eq:characteristic_atom}).
\end{enumerate}
We note that the backward-in-time simulation approach eliminates the need for ``aiming'' the secondary ENAs to the observer.

\section{Results of simulations and comparison with IBEX-Hi data} \label{sec:results}

\subsection{Contribution of different components}

\begin{figure*}
\includegraphics[width=\textwidth]{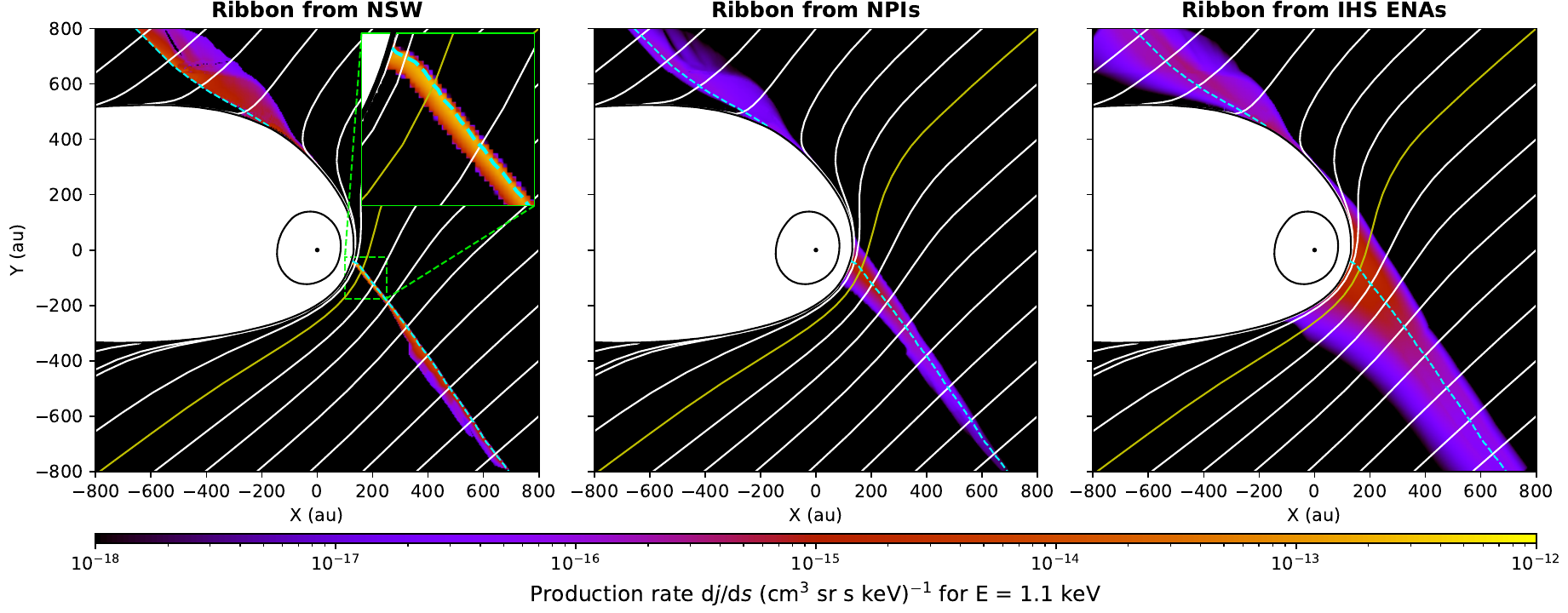}
\caption{
Spatial distribution of the ribbon production rate $\mathrm{d}j/\mathrm{d}s$ (cm$^3$ sr s keV)$^{-1}$ in the BV-plane for different components of the primary ENAs: Neutral Solar Wind atoms (first column), Neutralized Pickup Ions (second column), and ENAs from the inner heliosheath (third column). The production rate is calculated for the observer at the Sun, and energy $E$ = 1.1 keV. Black and white lines show the TS/HP boundaries and magnetic field lines, respectively. The cyan dashed lines show points where $\mathbf{r} \cdot \mathbf{B} = 0$. The yellow line is the magnetic field line that passes through the point (800 au, 800 au). The green inset box in the left panel shows the zoomed region [100 au, 250 au] $\times$ [-175 au, -25 au].
}
\label{fig:BV40_sources}
\end{figure*}

\begin{figure}
\includegraphics[width=\columnwidth]{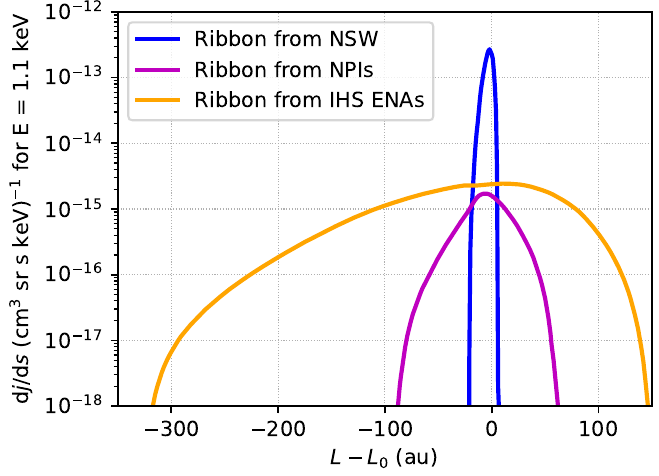}
\caption{
Spatial distribution of the ribbon production rate $\mathrm{d}j/\mathrm{d}s$ (cm$^3$ sr s keV)$^{-1}$ along the yellow magnetic field line in Figure \ref{fig:BV40_sources} as a function of the path length $L - L_0$ along the field line, where $L_0$ corresponds to the $\mathbf{r} \cdot \mathbf{B} = 0$ point. The production rate of secondary ENAs is shown for different components of the primary ENAs: NSW (blue line), NPIs (magenta line), and ENAs from the IHS (orange line).
}
\label{fig:production1D}
\end{figure}

First, let us analyze the spatial distribution of the ribbon sources outside the heliopause. Figure \ref{fig:BV40_sources} shows the ribbon production rate (or flux source function) in the BV-plane, which coincides with the HDP, for different components of the primary ENAs: Neutral Solar Wind atoms (Type 1.0), Neutralized Pickup Ions (Type 1.1), and ENAs from the inner heliosheath. The relation between the ribbon ENA flux $j$ and the production rate $(\mathrm{d}j/\mathrm{d}s)$ is the following: the flux is the integral of the production rate as a function of distance $s$ along the atom trajectory. The spatial source function of the ribbon ENA flux was calculated for the observer at the Sun and the specific energy $E$ = 1.1 keV (central energy of IBEX-Hi ESA3). The last column represents the sum of three components of the ribbon from the IHS ENAs (Types 2.0, 2.1, and 2.2). However, we note that only Type 2.1 has a non-negligible contribution to the fluxes at the energies under study (IBEX-Hi energy range), and Types 2.0 and 2.2 provide the fluxes at lower energy (see Figure \ref{fig:spectra} with 1D energy spectra below). Figure \ref{fig:production1D} shows the ribbon production rates as the functions of path length along the specific magnetic field line, which is indicated by the yellow line in Figure \ref{fig:BV40_sources}.

As can be seen from Figures \ref{fig:BV40_sources} and \ref{fig:production1D}, the production rate is not zero at the points where the radius vector is roughly perpendicular to the magnetic field line ($\mathbf{r} \cdot \mathbf{B} \approx 0$). The NSW component has the narrowest structure compared to NPIs and IHS ENAs components. We note that the width of the source region correlates with the size of the region where primary ENAs originate (supersonic solar wind region and inner heliosheath for the second and third columns, respectively). The qualitative analysis of the spatial distribution of the ribbon production rate is presented in \ref{app:spatial_sources}.

\begin{figure*}
\includegraphics[width=\textwidth]{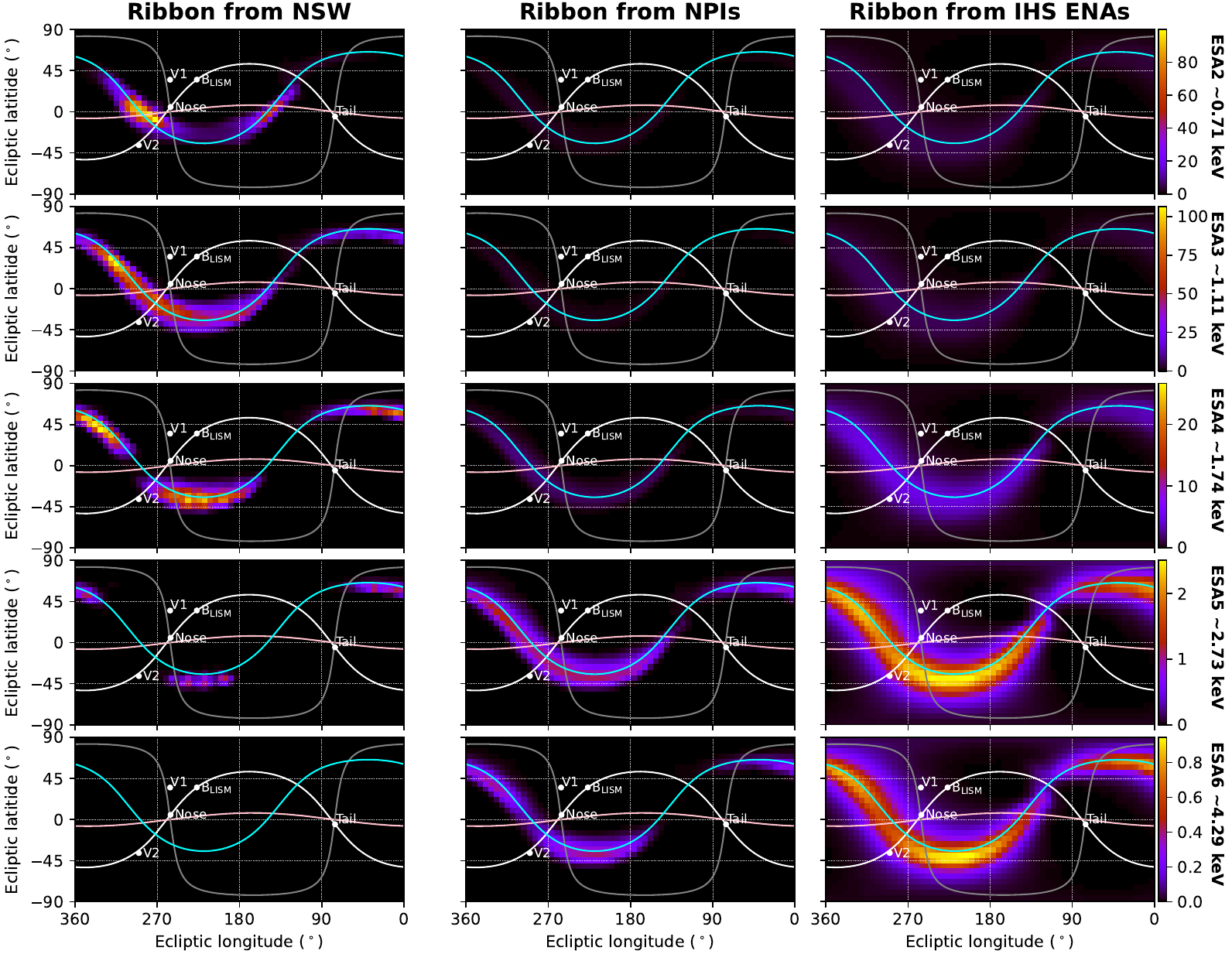}
\caption{
Simulated full-sky ribbon flux maps in ecliptic (J2000) coordinates for the IBEX-Hi energy channels 2–6 (by rows). The first, second, and third columns represent the contributions of ribbon H atoms that originate from the pickup protons in the outer heliosheath that are produced by charge exchange of Neutral Solar Wind atoms, Neutralized Pickup Ions, and ENAs from the inner heliosheath, respectively. The units of flux are (cm$^2$ sr s keV)$^{-1}$. Pink, gray, and white lines show the solar equatorial and polar (meridional) planes and the HDP. The cyan curve shows the IBEX ribbon directions based on the analysis of \citet{funsten2013}. The white dots with captions show principal directions.
}
\label{fig:BV40_components}
\end{figure*}

Now let us consider the distribution of the ribbon in the sky. Figure \ref{fig:BV40_components} shows the full-sky ENA flux maps in ecliptic coordinates as observed by IBEX-Hi with separate contributions of ribbon H atoms that originate from the pickup protons in the outer heliosheath that are produced in charge exchange of Neutral Solar Wind atoms (Type 1.0), Neutralized Pickup Ions (Type 1.1), and ENAs from the inner heliosheath (only Type 2.1; Types 2.0 and 2.2 have negligible contribution). For reference, we also plot the ribbon directions in the maps (cyan line) with the ribbon center at ($\lambda_{\rm RC}$, $\beta_{\rm RC}$) = (219.$^\circ$2, 39.$^\circ$9) in heliocentric ecliptic coordinates and a half cone angle of 74.$^\circ$5 estimated based on the circularity analysis by \citet{funsten2013}.

As can be seen from Figure \ref{fig:BV40_components}, the contribution of ribbon from NSW (first column) to the fluxes at energy channels $\sim$0.71, 1.11, and 1.74 keV is dominant. The fluxes at the lowest energy channel $\sim$0.71 keV are an imprint of the NSW produced from the slow ($\sim$400 km/s) solar wind close to the equatorial plane shown with the pink line. At the same time, the atoms that originate from the fast ($\sim$600-700 km/s) solar wind protons are responsible for the two high-latitude regions of fluxes seen at $\sim$1.74 keV (above and below the solar equatorial plane). 

NPIs and IHS ENAs (Type 2.1 in particular), as more energetic components than NSW, are the main contributors to the ribbon sources at the highest IBEX-Hi energy channels ($\sim$2.73 and $\sim$4.29 keV). We note that (a) the contribution of IHS ENAs to the total flux at these energies is even higher than that of NPIs, and (b) IHS ENAs produce a relatively wide flux structure in the sky, so this component must be considered if one studies the ribbon width. However, most of the previous studies neglect this component (primarily due to the difficulty of its modeling). 

To our knowledge, \citet{zirnstein2016a}, \citet{schwadron2019}, and \citet{gamayunov2017, gamayunov2019} are the only studies, which considered the contribution of IHS ENAs to the ribbon production. Still, in their simulations \citet{zirnstein2016a} and \citet{gamayunov2017, gamayunov2019} used a simplified approach assuming a kappa distribution of the plasma with constant $\kappa = 1.63$ throughout the IHS, while \citet{schwadron2019} used an approximation for the differential ENA flux from the inner heliosheath, assuming a position-independent source. Several studies also considered the contribution of IHS ENAs, but their analyses are limited to the stability of pickup proton distribution in the outer heliosphere \citep[e.g.,][]{roytershteyn2019, mousavi2025}. The distinctive feature of our model is that we consider the pickup proton velocity distribution kinetically, accounting for thermal and suprathermal proton populations in the heliosphere separately \citep{baliukin2020}. This allows us to simulate different populations of ENAs from the inner heliosheath (Types 2.0, 2.1, and 2.2).

The other thing worth mentioning is that at the highest energy channel, Types 1.1 and 2.1 also produce high-latitude maxima of fluxes (even though not as prominent as Type 1.0). The reason is that their parent pickup protons have a broader filled-shell distribution at high latitudes, inherited from the solar wind velocity profile.

\subsection{Comparison with IBEX-Hi data} \label{sec:comparison}

For the comparison with our model simulation, we use the IBEX team-validated ribbon/GDF separated flux maps presented by \citet{mccomas2024}. \href{https://ibex.princeton.edu/DataRelease18}{Data Release 18} provides not only the best guess (median) maps but also the maps with upper and lower reasonable values of ribbon and GDF fluxes. The yearly data is available from 2009 to 2022. However, we averaged the maps over only 11 years (2009 -- 2019) to compare with the simulation results of our stationary model that employs solar cycle averaged solar wind boundary conditions. For our analysis, we use ram observations with the survival probability and Compton-Getting corrections applied.

\begin{figure*}
\includegraphics[width=\textwidth]{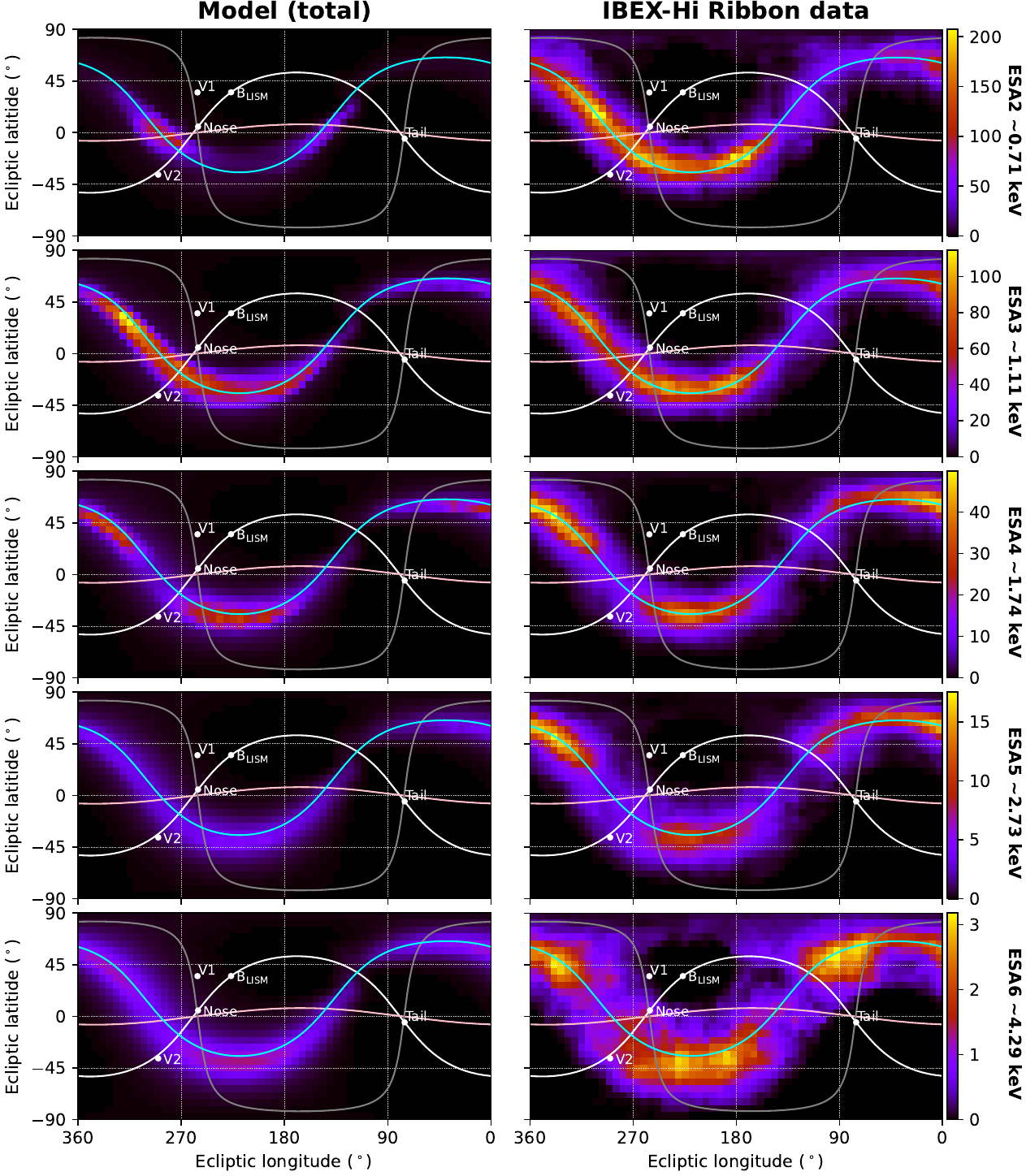}
\caption{
The caption is the same as for Figure \ref{fig:BV40_components}, but the left column presents the total model -- the sum of three components (ribbon ENAs produced from NSW, NPIs, IHS ENAs), while the right column shows the IBEX-Hi ribbon separated median maps (with Compton-Getting and survival probability corrections applied) from \citet{mccomas2024} averaged over 2009 -- 2019. 
}
\label{fig:BV40_total}
\end{figure*}

Figure \ref{fig:BV40_total} presents the total model (left column), the sum of three components shown in Figure \ref{fig:BV40_components} (ribbon ENAs produced from NSW, NPIs, and IHS ENAs), and the IBEX ribbon separated median maps. As can be seen in this figure, the model qualitatively replicates the ribbon shape well. However, the data provides a notably wider ribbon structure in the sky compared to the model. The reason for that is twofold. On the one hand, the simulations are based on the assumption of a complete absence of pitch-angle scattering in the outer heliosheath, which is an idealization. With some pitch-angle scattering, we expect to see a broader, more diffused flux structure in the sky. The simulations of the IBEX ribbon with realistic pitch-angle scattering efficiency will be done in future work. On the other hand, since the problem of the ribbon separation from the GDF is ill-posed (there is no unique solution), the ribbon width may be overestimated in the data.

As can be seen in Figure \ref{fig:BV40_total}, for the ribbon directions close to the solar equatorial plane (pink line), the model reproduces even the values of the fluxes. The most pronounced quantitative difference seen in the maps is the model deficit of fluxes in the high latitude regions (at heliolatitudes $\sim \pm 45^\circ$). This lack of fluxes in the model compared to the data is more pronounced at the lowest ($\sim$0.71 keV) and highest ($\sim$2.73 and $\sim$4.29 keV) energy channels. 

\begin{figure*}
\includegraphics[width=\textwidth]{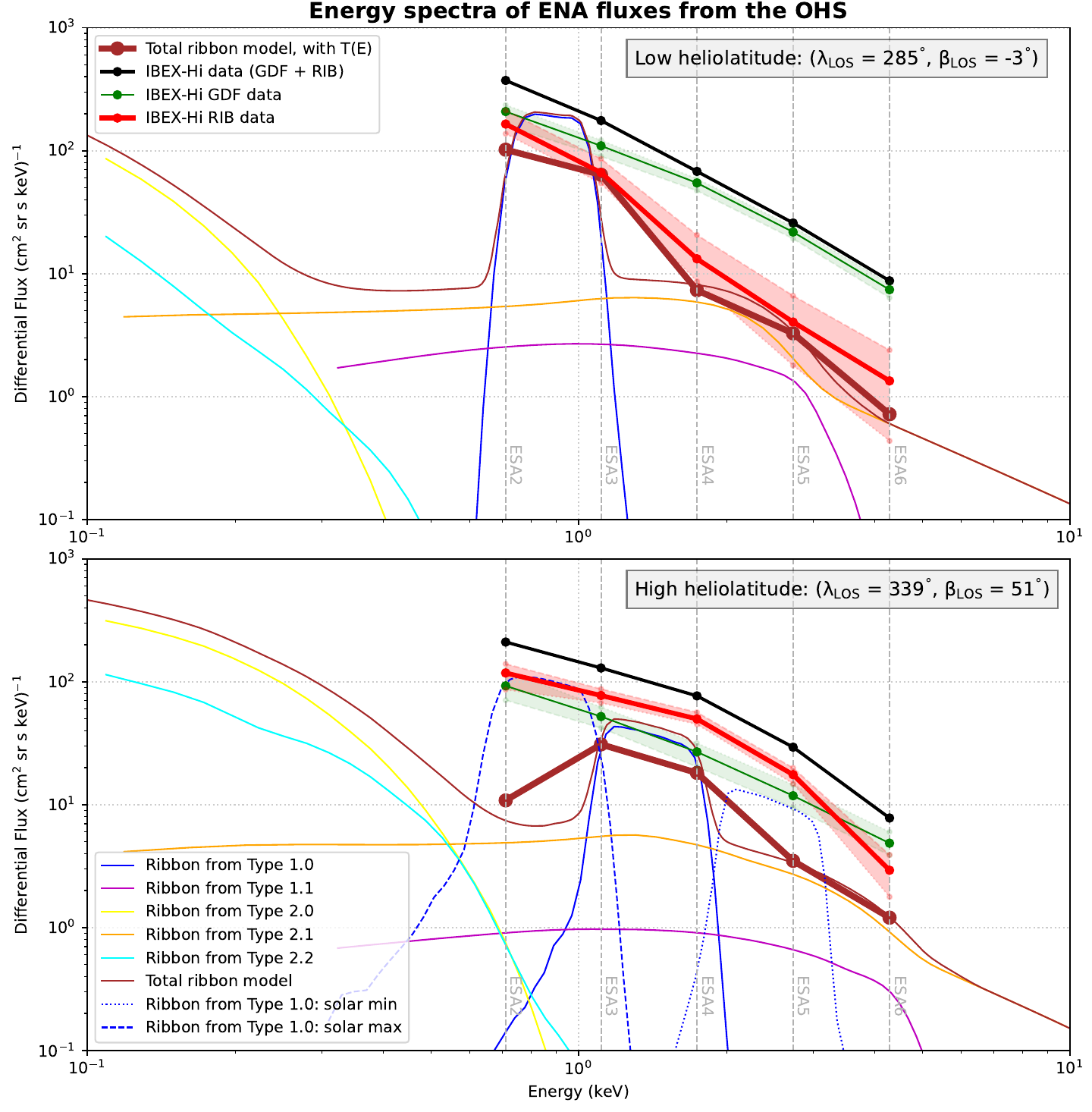}
\caption{
Energy spectra of ENA fluxes from the outer heliosheath in two directions -- within the solar equatorial plane (top panel) and towards the ribbon ``knot'' (bottom panel). The black lines show the initial IBEX-Hi data, while the green and red solid/dashed/dotted lines with dots show the GDF and ribbon median/upper/lower data, respectively. The blue, magenta, yellow, orange, and cyan solid lines show the ribbon produced from primary ENAs of Types 1.0, 1.1, 2.0, 2.1, and 2.2, respectively. The total model spectra are shown with the thin brown solid lines, while the brown dots represent the flux values at specific IBEX-Hi energy steps (calculated with energy transmission functions of the ESAs taken into account). The blue dotted and dashed lines show the ribbon ENAs spectra produced from primary ENAs of Type 1.0 calculated for typical solar minimum and maximum conditions, respectively (see text for the details).
}
\label{fig:spectra}
\end{figure*}

To study this issue in more detail, we performed the simulations of ENA energy spectra in two selected directions -- within the solar equatorial plane ($\lambda_{\rm LOS}$ = 285$^\circ$, $\beta_{\rm LOS}$ = -3$^\circ$) and towards the so-called ribbon ``knot'' with high heliolatitude ($\lambda_{\rm LOS}$ = 339$^\circ$, $\beta_{\rm LOS}$ = +51$^\circ$). The results of the simulations are shown in Figure \ref{fig:spectra}. As can be seen from the comparison of the model simulations with data, the primary ENAs of Types 2.0 and 2.2 (yellow and cyan lines) contribute to the ribbon spectra at energies lower than IBEX-Hi can observe, and, therefore, they can be neglected in the data analysis.

For the in-ecliptic direction (top panel), there is a good quantitative agreement -- most of the model fluxes calculated for IBEX-Hi energy steps (brown dots) fall within the lower and upper ribbon flux estimations (red shaded region). It is important to note that the Type 2.1 contribution (orange line) to the ENA fluxes at the highest energy steps ($\sim$1.74, $\sim$2.73, and $\sim$4.29 keV) is dominant, so it is not possible to explain the ribbon data without consideration of this component. We also note that the peak of the Type 1.0 ribbon component is relatively broad. It is the manifestation of the solar wind deceleration with distance (due to the charge exchange process) since secondary ENAs inherit the velocity distribution of primary NSW atoms.

For the high-latitude ``knot'' direction (bottom panel), the model compares poorly with the data -- the fluxes in the model are $\sim$2--10 times lower. The model fluxes peak at the energy steps $\sim$1.11 keV and $\sim$1.74 keV, where the ribbon ENAs from the NSW (Type 1.0) component dominates. From the comparison with the energy spectra towards the in-ecliptic direction, it is evident that this component is shifted to higher energies (due to the faster solar wind velocity at higher heliolatitude). The spectra of ribbon ENAs from Types 1.1 and 2.1 are shifted as well, following a broader filled-shell distribution of parent pickup protons at high heliolatitudes. 

We suggest the following explanation for the deficit of model fluxes at high latitudes for the lowest ($\sim$0.71 keV) and highest ($\sim$2.73 and $\sim$4.29 keV) energy channels. It is due to the essentially non-stationary behavior of the solar wind during the solar cycle, which is not captured by our stationary model simulations. The problem is that the stationary calculations with solar cycle averaged conditions for the solar wind parameters at 1 au are not equivalent to the time-dependent calculations that are then averaged over the solar cycle. At high heliolatitudes during the solar minimum periods, the solar wind is typically faster and less dense, and during solar maxima, it is slower and more dense compared to the solar cycle average values. Therefore, in the time-dependent simulations, we expect to see some effective broadening of the NSW energy spectrum (Type 1.0, blue solid curve) in the bottom panel of Figure \ref{fig:spectra}, which should narrow the gap between the model and data.

\begin{figure*}
\includegraphics[width=\textwidth]{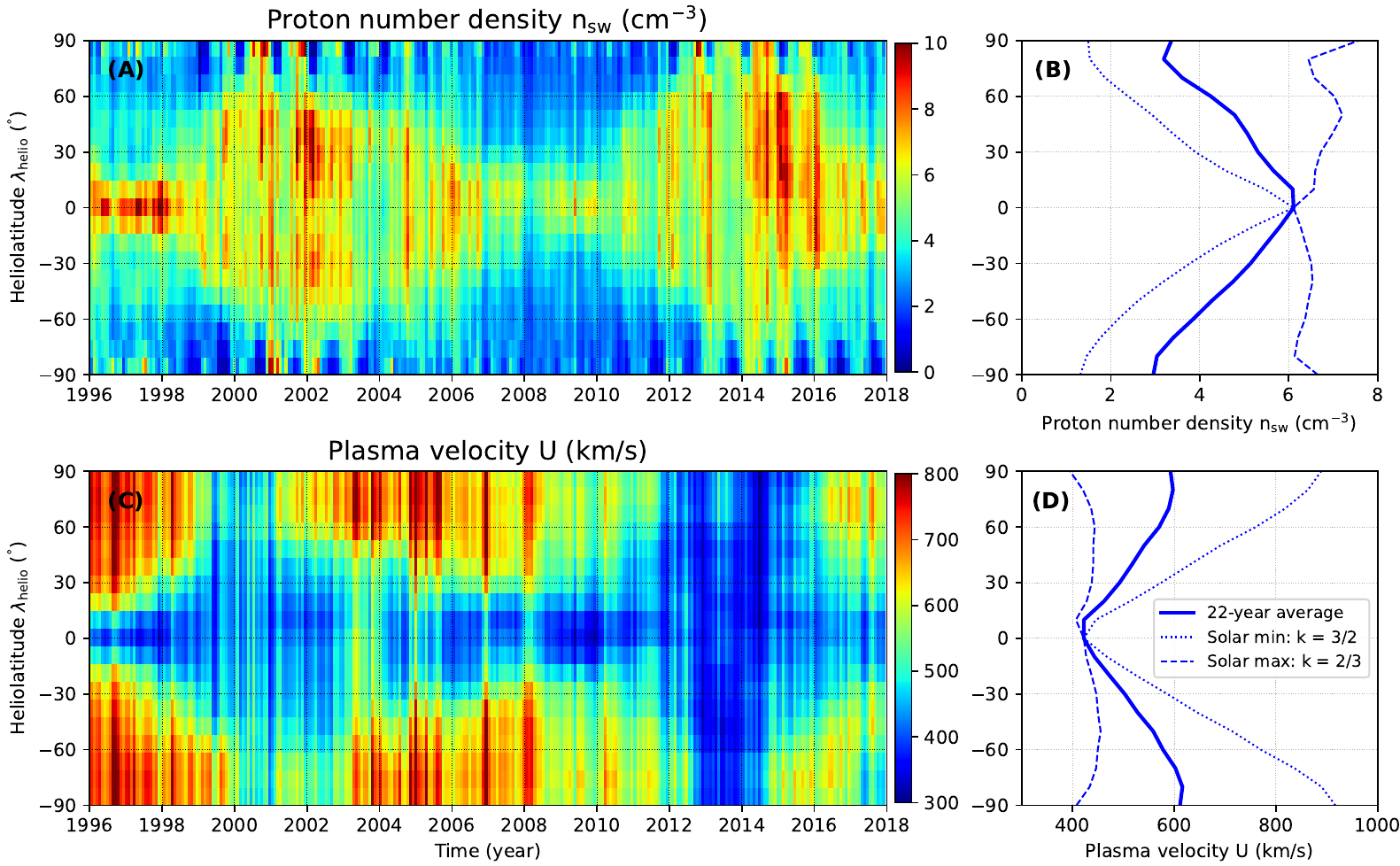}
\caption{
Solar wind proton number density (panel A) and velocity (panel C) at 1 au as functions of the heliolatitude and time. Panels (B) and (D) show the 22-year (1996 -- 2018) solar cycle averaged profiles (solid curves) and modified profiles typical for solar minimum and maximum conditions (dotted and dashed lines were calculated using $k = 3/2$ and $k = 2/3$, respectively; see text for details).
}
\label{fig:sw_profile}
\end{figure*}

Even though the time-dependent simulations are beyond the scope of this paper, as a ``proof of concept'', we provide some results of additional calculations with modified solar wind parameters. First, we introduce a coefficient \begin{equation}
    \chi = 1 + \left| \frac{2 \lambda_{\rm helio}}{\pi} \right| (k - 1),
\end{equation}
which is linearly dependent on the heliolatitude $\lambda_{\rm helio}$ and parameter $k$ that defines the value of scaling at the solar poles ($\lambda_{\rm helio} = \pm \pi/2$). Second, to calculate the NSW velocity distribution function (Type 1.0), we use modified thermal solar wind parameters in the supersonic solar wind region:
\begin{equation}
    \mathbf{U^*} = \mathbf{U} \chi, \: n_{\rm sw}^* = \frac{n_{\rm sw}}{\chi^2},
\end{equation}
where $\mathbf{U}$ and $n_{\rm sw}$ are plasma velocity and thermal solar wind number density from the simulations of the global model that utilizes the 22-year averaged solar wind parameters at 1 au. Note that the introduced modification conserves the dynamic pressure. Therefore, the $k > 1$ case corresponds to the increased solar wind velocity (typical for solar minima), while the $k < 1$ case represents conditions of the solar maxima. 

Figure \ref{fig:sw_profile} shows the solar wind proton number density (panel A) and velocity (panel C) at 1 AU as functions of the heliolatitude and time \citep[for details see Appendix A in][]{izmod2020}, and panels (B) and (D) show the 22-year (1996 -- 2018) averaged profiles (solid curves), as well as modified profiles typical for solar minimum and maximum conditions (dotted and dashed lines calculated using $k = 3/2$ and $k = 2/3$, respectively). In its turn, the blue dotted and dashed lines in the bottom panel of Figure \ref{fig:spectra} show the resulting NSW energy spectra calculated using the modified solar wind parameters (also assuming $k = 3/2$ and $k = 2/3$). Even though the calculations are not self-consistent, they show a significant variation of the ENA ribbon spectrum with the solar cycle and signify the importance of the time-dependent effects.

\section{Conclusions and discussion} \label{sec:conclusions}

In this work, we have developed a fully kinetic model that allows us to simulate the ENA fluxes from the OHS using the pre-calculated global distributions of plasma and H atoms in the heliosphere. The model uses a minimum number of assumptions, with the only critical one regarding the dynamics of the pickup protons in the OHS -- we assumed that there are no pitch angle scattering and energy diffusion outside the heliopause. Therefore, our work provides only the crucial first step for understanding the ribbon fluxes. The impact of these processes on the ribbon fluxes will be examined in future studies. The main conclusions of this work can be summarized as follows.

\begin{enumerate}

\item The model of ENAs from the outer heliosheath, which is based on the solution of the focused transport equation for the anisotropic velocity distribution function of pickup protons outside the heliopause under the scatter-free limit, reproduces the IBEX-Hi ribbon separated data qualitatively well. The model fluxes do not require specific scaling to fit the data (at least at low heliolatitudes). This result suggests that either the scattering remains weak or a spatial retention proposed by \citet{schwadron2013} is required, which is provided by strong scattering. However, the major issue of the spatial retention models is that simulated fluxes are smaller than the observed ones \citep[by a factor of $\sim$4; see, e.g.,][]{zirnstein2023}.

\item In the modeling, we considered all possible neutral components from the solar wind region and studied their relative contribution to the production of the secondary ENA fluxes from the outer heliosheath. We have shown that the NSW is responsible for the fluxes at the lowest IBEX-Hi energy channels ($\sim$0.71 keV, $\sim$1.11 keV), while the NPIs and ENAs from the IHS produce the observed fluxes at the highest IBEX-Hi energy channel ($\sim$4.29 keV). At low heliolatitudes, the fluxes at intermediate energy channels ($\sim$1.74 keV, $\sim$2.73 keV) are also defined by secondary ENA fluxes produced by NPIs and ENAs from the IHS. At high heliolatitudes, the fluxes at these energies are subject to the mutual contribution of all primary ENA components, and during the solar minima, the NSW can dominate even at $\sim$2.73 keV. The contribution of ENAs from the IHS to the ribbon production was considered in the frame of the kinetic model for the first time. The new finding is that this component is the main contributor to the total flux at the highest energy steps. We also note that it produces a relatively broad flux structure in the sky, so this component must be taken into account in the analysis of ribbon width, which is the indicator of the pitch-angle scattering efficiency.

\item The main difference between our model predictions and IBEX-Hi ribbon data is seen at high heliolatitudes, where the model produces significantly ($\sim$2--10 times) lower fluxes. The possible explanation is the essentially non-stationary solar wind behavior, which is not captured by the stationary simulations presented in this work. While our model utilizes solar-cycle averaged boundary conditions at 1 au, the solar wind shows significant variability in its parameters (number density and, especially, velocity) at high heliolatitudes during the solar cycle. In the non-stationary case, we expect to see a broader NSW energy spectrum, which should provide a better agreement between the model and data. However, the time-dependent model simulations are beyond the scope of this paper.

\end{enumerate}

The global model simulations of the plasma and H atom distribution in the heliosphere utilized in this work assume the following parameters in the LISM: $n_{\rm p,LISM} = 0.04$ cm$^{-3}$ and $n_{\rm H,LISM} = 0.14$ cm$^{-3}$ \citep[see the justification for such low values in Section 4 of][]{izmod2015}. However, recent Voyager 1/Plasma Wave Subsystem (PWS) observations of the electron density showed that starting from $\sim$149 au and up to 161.4 au, the density has remained relatively constant at 0.147 cm$^{-3}$ \citep{kurth2024}. \citet{fraternale2024} compared these data with global model predictions and described challenges in reproducing PWS observations associated with the presence of He$^+$ ions in the LISM in particular. Their study suggests that the proton number density in the LISM should exceed $\sim$0.07 cm$^{-3}$. Moreover, \citet{powell2024} showed that the model simulations with higher proton and neutral H number densities ($n_{\rm p,LISM} = 0.06$ cm$^{-3}$ and $n_{\rm H,LISM} = 0.18$ cm$^{-3}$) are in better agreement with Lyman-$\alpha$ absorption profiles towards nearby stars observed by the Hubble Space Telescope (note that their simulations did not account for He ions). The analysis of measurements made by the Solar Wind Around Pluto (SWAP) instrument on New Horizons spacecraft also suggests that the neutral H density in the LISM may be higher than the value used in our work \citep{swaczyna_etal:20a}. With the higher proton and neutral number density assumed in the LISM, one could expect higher ribbon fluxes in the model simulations (however, the effect is non-linear). In addition to applying a more realistic model with non-zero pitch-angle scattering, which redistributes the fluxes, making the ribbon structure more diffused and broader in the sky, we expect to see better agreement with IBEX observations. We leave the development of the improved model with more realistic pickup proton dynamics beyond the heliopause for future studies.

We note that the description of the ribbon model used in this work is made in the generalized time-dependent form with time and its derivatives preserved in all the equations, so the model can be easily used for the non-stationary simulations. However, to run non-stationary simulations, one should first obtain time-dependent global plasma and H atom distributions in the heliosphere. In future studies, we also plan to utilize the time-dependent version of the SW/LISM interaction model by \citet{izmod2015, izmod2020} to produce such distributions and study the time variability of the IBEX ribbon.

It is important to note that all the simulations presented in this paper have been made using the pre-calculated global distributions obtained within the heliosphere model by \citet{izmod2015, izmod2020}, which treats all the charged particles as a single fluid. So, the simulations presented in this work are not self-consistent. However, our preliminary simulations of a 3D self-consistent multi-component model, which considers the population of pickup protons kinetically, suggest that changes in the global structure of the heliosphere are quite small (compared to the single fluid case). Therefore, all the results and conclusions obtained in the frame of our non-self-consistent model could be considered valid. The results of global simulations of the heliosphere using the multi-component model will be presented in the forthcoming study.

In summary, the IBEX ribbon remains a focal point of heliospheric research, driving advancements in both observational techniques and theoretical modeling. Continued efforts in this field promise to deepen our understanding of the complex interactions at the heliospheric boundary. Ongoing and future missions, such as the Interstellar Mapping and Acceleration Probe \citep[IMAP;][]{mccomas2018}, are expected to provide higher resolution data and more comprehensive measurements, which is crucial for refining existing models and potentially uncovering new mechanisms behind the ribbon formation.

\section*{Acknowledgments}
The work is funded by the Chinese Academy of Sciences President’s International Fellowship Initiative Grant No. 2025PVB0060. IB thanks the International Space Science Institute -- Beijing (ISSI-BJ) for providing the work facilities and Committee on Space Research (COSPAR) for travel support for the 45th COSPAR Scientific Assembly, where the results have been presented and discussed. The authors thank D. B. Alexashov for providing the global distributions of plasma and hydrogen atoms in the heliosphere, S. D. Korolkov for fruitful discussions, and the IBEX team for preparing and making data available. This work also benefited from discussions and results enabled by the \href{https://shielddrivecenter.com/}{SHIELD DRIVE center}. The authors thank the reviewers for their valuable comments and help in improving the paper.

\bibliographystyle{jasr-model5-names}
\biboptions{authoryear}
\bibliography{bibliography}

\begin{thebibliography}{73}
\expandafter\ifx\csname natexlab\endcsname\relax\def\natexlab#1{#1}\fi
\ifx\xfnm\relax \def\xfnm[#1]{\unskip,\space#1}\fi

\bibitem[{{Baliukin} et~al.(2020){Baliukin}, {Izmodenov} \& {Alexashov}}]{baliukin2020}
\bibinfo{author}{{Baliukin}, I.~I.}, \bibinfo{author}{{Izmodenov}, V.~V.},  \& \bibinfo{author}{{Alexashov}, D.~B.} (\bibinfo{year}{2020}).
\newblock \bibinfo{title}{{Heliospheric energetic neutral atoms: Non-stationary modelling and comparison with IBEX-Hi data}}.
\newblock {\it \bibinfo{journal}{\mnras}\/},  {\it \bibinfo{volume}{499}\/}\bibinfo{issue}{(1)}, \bibinfo{pages}{441--454}. \DOIprefix\doi{10.1093/mnras/staa2862}. \href{http://arxiv.org/abs/2009.14062}{\tt arXiv:2009.14062}.

\bibitem[{{Baliukin} et~al.(2022){Baliukin}, {Izmodenov} \& {Alexashov}}]{baliukin2022}
\bibinfo{author}{{Baliukin}, I.~I.}, \bibinfo{author}{{Izmodenov}, V.~V.},  \& \bibinfo{author}{{Alexashov}, D.~B.} (\bibinfo{year}{2022}).
\newblock \bibinfo{title}{{Energetic pickup proton population downstream of the termination shock as revealed by IBEX-Hi data}}.
\newblock {\it \bibinfo{journal}{\mnras}\/},  {\it \bibinfo{volume}{509}\/}\bibinfo{issue}{(4)}, \bibinfo{pages}{5437--5453}. \DOIprefix\doi{10.1093/mnras/stab3214}. \href{http://arxiv.org/abs/2110.15930}{\tt arXiv:2110.15930}.

\bibitem[{{Baliukin} et~al.(2023){Baliukin}, {Izmodenov} \& {Alexashov}}]{baliukin2023}
\bibinfo{author}{{Baliukin}, I.~I.}, \bibinfo{author}{{Izmodenov}, V.~V.},  \& \bibinfo{author}{{Alexashov}, D.~B.} (\bibinfo{year}{2023}).
\newblock \bibinfo{title}{{Adiabatic energy change in the inner heliosheath: how does it affect the distribution of pickup protons and energetic neutral atom fluxes?}}
\newblock {\it \bibinfo{journal}{\mnras}\/},  {\it \bibinfo{volume}{525}\/}\bibinfo{issue}{(3)}, \bibinfo{pages}{3281--3286}. \DOIprefix\doi{10.1093/mnras/stad2518}. \href{http://arxiv.org/abs/2308.09145}{\tt arXiv:2308.09145}.

\bibitem[{{Chalov}(2006)}]{chalov2006}
\bibinfo{author}{{Chalov}, S.} (\bibinfo{year}{2006}).
\newblock \bibinfo{title}{{Interstellar Pickup Ions and Injection Problem for Anomalous Cosmic Rays: Theoretical Aspect}}.
\newblock {\it \bibinfo{journal}{ISSI Scientific Reports Series}\/},  {\it \bibinfo{volume}{5}\/}, \bibinfo{pages}{245--282}.

\bibitem[{{Chalov} et~al.(2010){Chalov}, {Alexashov}, {McComas}, {Izmodenov}, {Malama} \& {Schwadron}}]{chalov2010}
\bibinfo{author}{{Chalov}, S.~V.}, \bibinfo{author}{{Alexashov}, D.~B.}, \bibinfo{author}{{McComas}, D.} et~al. (\bibinfo{year}{2010}).
\newblock \bibinfo{title}{{Scatter-free Pickup Ions beyond the Heliopause as a Model for the Interstellar Boundary Explorer Ribbon}}.
\newblock {\it \bibinfo{journal}{\apjl}\/},  {\it \bibinfo{volume}{716}\/}\bibinfo{issue}{(2)}, \bibinfo{pages}{L99--L102}. \DOIprefix\doi{10.1088/2041-8205/716/2/L99}. \href{http://arxiv.org/abs/1003.4826}{\tt arXiv:1003.4826}.

\bibitem[{{Florinski} et~al.(2024){Florinski}, {Guzman}, {Kleimann}, {Baliukin}, {Ghanbari}, {Turner}, {Zieger}, {K{\'o}ta}, {Opher}, {Izmodenov}, {Alexashov}, {Giacalone} \& {Richardson}}]{florinski2024}
\bibinfo{author}{{Florinski}, V.}, \bibinfo{author}{{Guzman}, J.~A.}, \bibinfo{author}{{Kleimann}, J.} et~al. (\bibinfo{year}{2024}).
\newblock \bibinfo{title}{{Magnetic Trapping of Galactic Cosmic Rays in the Outer Heliosheath and Their Preferential Entry into the Heliosphere}}.
\newblock {\it \bibinfo{journal}{\apj}\/},  {\it \bibinfo{volume}{961}\/}\bibinfo{issue}{(2)}, \bibinfo{pages}{244}. \DOIprefix\doi{10.3847/1538-4357/ad0b15}.

\bibitem[{{Florinski} \& {Heerikhuisen}(2017)}]{florinski2017}
\bibinfo{author}{{Florinski}, V.},  \& \bibinfo{author}{{Heerikhuisen}, J.} (\bibinfo{year}{2017}).
\newblock \bibinfo{title}{{Kinetic Properties of the Neutral Solar Wind}}.
\newblock {\it \bibinfo{journal}{\apj}\/},  {\it \bibinfo{volume}{838}\/}\bibinfo{issue}{(1)}, \bibinfo{pages}{50}. \DOIprefix\doi{10.3847/1538-4357/aa6441}.

\bibitem[{{Florinski} et~al.(2010){Florinski}, {Zank}, {Heerikhuisen}, {Hu} \& {Khazanov}}]{florinski2010}
\bibinfo{author}{{Florinski}, V.}, \bibinfo{author}{{Zank}, G.~P.}, \bibinfo{author}{{Heerikhuisen}, J.} et~al. (\bibinfo{year}{2010}).
\newblock \bibinfo{title}{{Stability of a Pickup Ion Ring-beam Population in the Outer Heliosheath: Implications for the IBEX Ribbon}}.
\newblock {\it \bibinfo{journal}{\apj}\/},  {\it \bibinfo{volume}{719}\/}\bibinfo{issue}{(2)}, \bibinfo{pages}{1097--1103}. \DOIprefix\doi{10.1088/0004-637X/719/2/1097}.

\bibitem[{{Fraternale} et~al.(2024){Fraternale}, {Pogorelov} \& {Bera}}]{fraternale2024}
\bibinfo{author}{{Fraternale}, F.}, \bibinfo{author}{{Pogorelov}, N.~V.},  \& \bibinfo{author}{{Bera}, R.~K.} (\bibinfo{year}{2024}).
\newblock \bibinfo{title}{{Constraining the Properties of the Multicomponent Local Interstellar Medium: MHD-kinetic Modeling Validated by Voyager and New Horizons Data}}.
\newblock {\it \bibinfo{journal}{\apjl}\/},  {\it \bibinfo{volume}{974}\/}\bibinfo{issue}{(1)}, \bibinfo{pages}{L15}. \DOIprefix\doi{10.3847/2041-8213/ad7e1c}.

\bibitem[{{Funsten} et~al.(2009){Funsten}, {Allegrini}, {Bochsler}, {Dunn}, {Ellis}, {Everett}, {Fagan}, {Fuselier}, {Granoff}, {Gruntman}, {Guthrie}, {Hanley}, {Harper}, {Heirtzler}, {Janzen}, {Kihara}, {King}, {Kucharek}, {Manzo}, {Maple}, {Mashburn}, {McComas}, {Moebius}, {Nolin}, {Piazza}, {Pope}, {Reisenfeld}, {Rodriguez}, {Roelof}, {Saul}, {Turco}, {Valek}, {Weidner}, {Wurz} \& {Zaffke}}]{funsten2009}
\bibinfo{author}{{Funsten}, H.~O.}, \bibinfo{author}{{Allegrini}, F.}, \bibinfo{author}{{Bochsler}, P.} et~al. (\bibinfo{year}{2009}).
\newblock \bibinfo{title}{{The Interstellar Boundary Explorer High Energy (IBEX-Hi) Neutral Atom Imager}}.
\newblock {\it \bibinfo{journal}{\ssr}\/},  {\it \bibinfo{volume}{146}\/}\bibinfo{issue}{(1-4)}, \bibinfo{pages}{75--103}. \DOIprefix\doi{10.1007/s11214-009-9504-y}.

\bibitem[{{Funsten} et~al.(2013){Funsten}, {DeMajistre}, {Frisch}, {Heerikhuisen}, {Higdon}, {Janzen}, {Larsen}, {Livadiotis}, {McComas}, {M{\"o}bius}, {Reese}, {Reisenfeld}, {Schwadron} \& {Zirnstein}}]{funsten2013}
\bibinfo{author}{{Funsten}, H.~O.}, \bibinfo{author}{{DeMajistre}, R.}, \bibinfo{author}{{Frisch}, P.~C.} et~al. (\bibinfo{year}{2013}).
\newblock \bibinfo{title}{{Circularity of the Interstellar Boundary Explorer Ribbon of Enhanced Energetic Neutral Atom (ENA) Flux}}.
\newblock {\it \bibinfo{journal}{\apj}\/},  {\it \bibinfo{volume}{776}\/}\bibinfo{issue}{(1)}, \bibinfo{pages}{30}. \DOIprefix\doi{10.1088/0004-637X/776/1/30}.

\bibitem[{{Fuselier} et~al.(2009){Fuselier}, {Bochsler}, {Chornay}, {Clark}, {Crew}, {Dunn}, {Ellis}, {Friedmann}, {Funsten}, {Ghielmetti}, {Googins}, {Granoff}, {Hamilton}, {Hanley}, {Heirtzler}, {Hertzberg}, {Isaac}, {King}, {Knauss}, {Kucharek}, {Kudirka}, {Livi}, {Lobell}, {Longworth}, {Mashburn}, {McComas}, {M{\"o}bius}, {Moore}, {Moore}, {Nemanich}, {Nolin}, {O'Neal}, {Piazza}, {Peterson}, {Pope}, {Rosmarynowski}, {Saul}, {Scherrer}, {Scheer}, {Schlemm}, {Schwadron}, {Tillier}, {Turco}, {Tyler}, {Vosbury}, {Wieser}, {Wurz} \& {Zaffke}}]{fuselier2009}
\bibinfo{author}{{Fuselier}, S.~A.}, \bibinfo{author}{{Bochsler}, P.}, \bibinfo{author}{{Chornay}, D.} et~al. (\bibinfo{year}{2009}).
\newblock \bibinfo{title}{{The IBEX-Lo Sensor}}.
\newblock {\it \bibinfo{journal}{\ssr}\/},  {\it \bibinfo{volume}{146}\/}\bibinfo{issue}{(1-4)}, \bibinfo{pages}{117--147}. \DOIprefix\doi{10.1007/s11214-009-9495-8}.

\bibitem[{{Galli} et~al.(2022){Galli}, {Baliukin}, {Bzowski}, {Izmodenov}, {Kornbleuth}, {Kucharek}, {M{\"o}bius}, {Opher}, {Reisenfeld}, {Schwadron} \& {Swaczyna}}]{galli2022}
\bibinfo{author}{{Galli}, A.}, \bibinfo{author}{{Baliukin}, I.~I.}, \bibinfo{author}{{Bzowski}, M.} et~al. (\bibinfo{year}{2022}).
\newblock \bibinfo{title}{{The Heliosphere and Local Interstellar Medium from Neutral Atom Observations at Energies Below 10 keV}}.
\newblock {\it \bibinfo{journal}{\ssr}\/},  {\it \bibinfo{volume}{218}\/}\bibinfo{issue}{(4)}, \bibinfo{pages}{31}. \DOIprefix\doi{10.1007/s11214-022-00901-7}.

\bibitem[{{Gamayunov} et~al.(2017){Gamayunov}, {Heerikhuisen} \& {Rassoul}}]{gamayunov2017}
\bibinfo{author}{{Gamayunov}, K.~V.}, \bibinfo{author}{{Heerikhuisen}, J.},  \& \bibinfo{author}{{Rassoul}, H.} (\bibinfo{year}{2017}).
\newblock \bibinfo{title}{{A Test of the Interstellar Boundary EXplorer Ribbon Formation in the Outer Heliosheath}}.
\newblock {\it \bibinfo{journal}{\apj}\/},  {\it \bibinfo{volume}{845}\/}\bibinfo{issue}{(1)}, \bibinfo{pages}{63}. \DOIprefix\doi{10.3847/1538-4357/aa7f70}.

\bibitem[{{Gamayunov} et~al.(2019){Gamayunov}, {Heerikhuisen} \& {Rassoul}}]{gamayunov2019}
\bibinfo{author}{{Gamayunov}, K.~V.}, \bibinfo{author}{{Heerikhuisen}, J.},  \& \bibinfo{author}{{Rassoul}, H.~K.} (\bibinfo{year}{2019}).
\newblock \bibinfo{title}{{Effect of the Interstellar Magnetic Field Draping around the Heliopause on the IBEX Ribbon}}.
\newblock {\it \bibinfo{journal}{\apjl}\/},  {\it \bibinfo{volume}{876}\/}\bibinfo{issue}{(2)}, \bibinfo{pages}{L21}. \DOIprefix\doi{10.3847/2041-8213/ab1b4f}.

\bibitem[{{Gazis} et~al.(1994){Gazis}, {Barnes}, {Mihalov} \& {Lazarus}}]{gazis1994}
\bibinfo{author}{{Gazis}, P.~R.}, \bibinfo{author}{{Barnes}, A.}, \bibinfo{author}{{Mihalov}, J.~D.} et~al. (\bibinfo{year}{1994}).
\newblock \bibinfo{title}{{Solar wind velocity and temperature in the outer heliosphere}}.
\newblock {\it \bibinfo{journal}{\jgr}\/},  {\it \bibinfo{volume}{99}\/}\bibinfo{issue}{(A4)}, \bibinfo{pages}{6561--6574}. \DOIprefix\doi{10.1029/93JA03144}.

\bibitem[{{Heerikhuisen} et~al.(2016){Heerikhuisen}, {Gamayunov}, {Zirnstein} \& {Pogorelov}}]{heerikhuisen2016}
\bibinfo{author}{{Heerikhuisen}, J.}, \bibinfo{author}{{Gamayunov}, K.~V.}, \bibinfo{author}{{Zirnstein}, E.~J.} et~al. (\bibinfo{year}{2016}).
\newblock \bibinfo{title}{{Neutral Atom Properties in the Direction of the IBEX Ribbon}}.
\newblock {\it \bibinfo{journal}{\apj}\/},  {\it \bibinfo{volume}{831}\/}\bibinfo{issue}{(2)}, \bibinfo{pages}{137}. \DOIprefix\doi{10.3847/0004-637X/831/2/137}.

\bibitem[{{Heerikhuisen} et~al.(2010){Heerikhuisen}, {Pogorelov}, {Zank}, {Crew}, {Frisch}, {Funsten}, {Janzen}, {McComas}, {Reisenfeld} \& {Schwadron}}]{heerikhuisen2010}
\bibinfo{author}{{Heerikhuisen}, J.}, \bibinfo{author}{{Pogorelov}, N.~V.}, \bibinfo{author}{{Zank}, G.~P.} et~al. (\bibinfo{year}{2010}).
\newblock \bibinfo{title}{{Pick-Up Ions in the Outer Heliosheath: A Possible Mechanism for the Interstellar Boundary EXplorer Ribbon}}.
\newblock {\it \bibinfo{journal}{\apjl}\/},  {\it \bibinfo{volume}{708}\/}\bibinfo{issue}{(2)}, \bibinfo{pages}{L126--L130}. \DOIprefix\doi{10.1088/2041-8205/708/2/L126}.

\bibitem[{{Isenberg}(1997)}]{isenberg1997}
\bibinfo{author}{{Isenberg}, P.~A.} (\bibinfo{year}{1997}).
\newblock \bibinfo{title}{{A hemispherical model of anisotropic interstellar pickup ions}}.
\newblock {\it \bibinfo{journal}{\jgr}\/},  {\it \bibinfo{volume}{102}\/}\bibinfo{issue}{(A3)}, \bibinfo{pages}{4719--4724}. \DOIprefix\doi{10.1029/96JA03671}.

\bibitem[{{Isenberg}(2014)}]{isenberg2014}
\bibinfo{author}{{Isenberg}, P.~A.} (\bibinfo{year}{2014}).
\newblock \bibinfo{title}{{Spatial Confinement of the IBEX Ribbon: A Dominant Turbulence Mechanism}}.
\newblock {\it \bibinfo{journal}{\apj}\/},  {\it \bibinfo{volume}{787}\/}\bibinfo{issue}{(1)}, \bibinfo{pages}{76}. \DOIprefix\doi{10.1088/0004-637X/787/1/76}. \href{http://arxiv.org/abs/1404.2170}{\tt arXiv:1404.2170}.

\bibitem[{{Isenberg} et~al.(2003){Isenberg}, {Smith} \& {Matthaeus}}]{isenberg2003}
\bibinfo{author}{{Isenberg}, P.~A.}, \bibinfo{author}{{Smith}, C.~W.},  \& \bibinfo{author}{{Matthaeus}, W.~H.} (\bibinfo{year}{2003}).
\newblock \bibinfo{title}{{Turbulent Heating of the Distant Solar Wind by Interstellar Pickup Protons}}.
\newblock {\it \bibinfo{journal}{\apj}\/},  {\it \bibinfo{volume}{592}\/}\bibinfo{issue}{(1)}, \bibinfo{pages}{564--573}. \DOIprefix\doi{10.1086/375584}.

\bibitem[{{Izmodenov} et~al.(2005){Izmodenov}, {Alexashov} \& {Myasnikov}}]{izmod2005b}
\bibinfo{author}{{Izmodenov}, V.}, \bibinfo{author}{{Alexashov}, D.},  \& \bibinfo{author}{{Myasnikov}, A.} (\bibinfo{year}{2005}).
\newblock \bibinfo{title}{{Direction of the interstellar H atom inflow in the heliosphere: Role of the interstellar magnetic field}}.
\newblock {\it \bibinfo{journal}{\aap}\/},  {\it \bibinfo{volume}{437}\/}\bibinfo{issue}{(3)}, \bibinfo{pages}{L35--L38}. \DOIprefix\doi{10.1051/0004-6361:200500132}.

\bibitem[{{Izmodenov}(2000)}]{izmod2000}
\bibinfo{author}{{Izmodenov}, V.~V.} (\bibinfo{year}{2000}).
\newblock \bibinfo{title}{{Physics and Gasdynamics of the Heliospheric Interface}}.
\newblock {\it \bibinfo{journal}{\apss}\/},  {\it \bibinfo{volume}{274}\/}, \bibinfo{pages}{55--69}. \DOIprefix\doi{10.1023/A:1026579418955}.

\bibitem[{{Izmodenov} \& {Alexashov}(2015)}]{izmod2015}
\bibinfo{author}{{Izmodenov}, V.~V.},  \& \bibinfo{author}{{Alexashov}, D.~B.} (\bibinfo{year}{2015}).
\newblock \bibinfo{title}{{Three-dimensional Kinetic-MHD Model of the Global Heliosphere with the Heliopause-surface Fitting}}.
\newblock {\it \bibinfo{journal}{\apjs}\/},  {\it \bibinfo{volume}{220}\/}\bibinfo{issue}{(2)}, \bibinfo{pages}{32}. \DOIprefix\doi{10.1088/0067-0049/220/2/32}. \href{http://arxiv.org/abs/1509.08685}{\tt arXiv:1509.08685}.

\bibitem[{{Izmodenov} \& {Alexashov}(2020)}]{izmod2020}
\bibinfo{author}{{Izmodenov}, V.~V.},  \& \bibinfo{author}{{Alexashov}, D.~B.} (\bibinfo{year}{2020}).
\newblock \bibinfo{title}{{Magnitude and direction of the local interstellar magnetic field inferred from Voyager 1 and 2 interstellar data and global heliospheric model}}.
\newblock {\it \bibinfo{journal}{\aap}\/},  {\it \bibinfo{volume}{633}\/}, \bibinfo{pages}{L12}. \DOIprefix\doi{10.1051/0004-6361/201937058}. \href{http://arxiv.org/abs/2001.03061}{\tt arXiv:2001.03061}.

\bibitem[{{Izmodenov} et~al.(2001){Izmodenov}, {Gruntman} \& {Malama}}]{izmod2001}
\bibinfo{author}{{Izmodenov}, V.~V.}, \bibinfo{author}{{Gruntman}, M.},  \& \bibinfo{author}{{Malama}, Y.~G.} (\bibinfo{year}{2001}).
\newblock \bibinfo{title}{{Interstellar hydrogen atom distribution function in the outer heliosphere}}.
\newblock {\it \bibinfo{journal}{\jgr}\/},  {\it \bibinfo{volume}{106}\/}\bibinfo{issue}{(A6)}, \bibinfo{pages}{10681--10690}. \DOIprefix\doi{10.1029/2000JA000273}.

\bibitem[{{Izmodenov} et~al.(2009){Izmodenov}, {Malama}, {Ruderman}, {Chalov}, {Alexashov}, {Katushkina} \& {Provornikova}}]{izmod2009}
\bibinfo{author}{{Izmodenov}, V.~V.}, \bibinfo{author}{{Malama}, Y.~G.}, \bibinfo{author}{{Ruderman}, M.~S.} et~al. (\bibinfo{year}{2009}).
\newblock \bibinfo{title}{{Kinetic-Gasdynamic Modeling of the Heliospheric Interface}}.
\newblock {\it \bibinfo{journal}{\ssr}\/},  {\it \bibinfo{volume}{146}\/}\bibinfo{issue}{(1-4)}, \bibinfo{pages}{329--351}. \DOIprefix\doi{10.1007/s11214-009-9528-3}.

\bibitem[{{Katushkina} et~al.(2015){Katushkina}, {Izmodenov}, {Alexashov}, {Schwadron} \& {McComas}}]{katushkina2015a}
\bibinfo{author}{{Katushkina}, O.~A.}, \bibinfo{author}{{Izmodenov}, V.~V.}, \bibinfo{author}{{Alexashov}, D.~B.} et~al. (\bibinfo{year}{2015}).
\newblock \bibinfo{title}{{Interstellar Hydrogen Fluxes Measured by IBEX-Lo in 2009: Numerical Modeling and Comparison with the Data}}.
\newblock {\it \bibinfo{journal}{\apjs}\/},  {\it \bibinfo{volume}{220}\/}\bibinfo{issue}{(2)}, \bibinfo{pages}{33}. \DOIprefix\doi{10.1088/0067-0049/220/2/33}. \href{http://arxiv.org/abs/1509.08754}{\tt arXiv:1509.08754}.

\bibitem[{{Korolkov} \& {Izmodenov}(2022)}]{korolkov2022}
\bibinfo{author}{{Korolkov}, S.~D.},  \& \bibinfo{author}{{Izmodenov}, V.~V.} (\bibinfo{year}{2022}).
\newblock \bibinfo{title}{{Shock-wave heating mechanism of the distant solar wind: Explanation of Voyager-2 data}}.
\newblock {\it \bibinfo{journal}{\aap}\/},  {\it \bibinfo{volume}{667}\/}, \bibinfo{pages}{L5}. \DOIprefix\doi{10.1051/0004-6361/202244523}. \href{http://arxiv.org/abs/2210.15032}{\tt arXiv:2210.15032}.

\bibitem[{{Kowalska-Leszczynska} et~al.(2020){Kowalska-Leszczynska}, {Bzowski}, {Kubiak} \& {Sok{\'o}{\l}}}]{kowalska2020}
\bibinfo{author}{{Kowalska-Leszczynska}, I.}, \bibinfo{author}{{Bzowski}, M.}, \bibinfo{author}{{Kubiak}, M.~A.} et~al. (\bibinfo{year}{2020}).
\newblock \bibinfo{title}{{Update of the Solar Ly{\ensuremath{\alpha}} Profile Line Model}}.
\newblock {\it \bibinfo{journal}{\apjs}\/},  {\it \bibinfo{volume}{247}\/}\bibinfo{issue}{(2)}, \bibinfo{pages}{62}. \DOIprefix\doi{10.3847/1538-4365/ab7b77}. \href{http://arxiv.org/abs/2001.07065}{\tt arXiv:2001.07065}.

\bibitem[{{Kurth}(2024)}]{kurth2024}
\bibinfo{author}{{Kurth}, W.~S.} (\bibinfo{year}{2024}).
\newblock \bibinfo{title}{{Voyager 1 Electron Densities in the Very Local Interstellar Medium to beyond 160 au}}.
\newblock {\it \bibinfo{journal}{\apjl}\/},  {\it \bibinfo{volume}{963}\/}\bibinfo{issue}{(1)}, \bibinfo{pages}{L6}. \DOIprefix\doi{10.3847/2041-8213/ad2617}.

\bibitem[{{Lallement} et~al.(2005){Lallement}, {Qu{\'e}merais}, {Bertaux}, {Ferron}, {Koutroumpa} \& {Pellinen}}]{lallement2005}
\bibinfo{author}{{Lallement}, R.}, \bibinfo{author}{{Qu{\'e}merais}, E.}, \bibinfo{author}{{Bertaux}, J.~L.} et~al. (\bibinfo{year}{2005}).
\newblock \bibinfo{title}{{Deflection of the Interstellar Neutral Hydrogen Flow Across the Heliospheric Interface}}.
\newblock {\it \bibinfo{journal}{Science}\/},  {\it \bibinfo{volume}{307}\/}\bibinfo{issue}{(5714)}, \bibinfo{pages}{1447--1449}. \DOIprefix\doi{10.1126/science.1107953}.

\bibitem[{{Lallement} et~al.(2010){Lallement}, {Qu{\'e}merais}, {Koutroumpa}, {Bertaux}, {Ferron}, {Schmidt} \& {Lamy}}]{lallement2010}
\bibinfo{author}{{Lallement}, R.}, \bibinfo{author}{{Qu{\'e}merais}, E.}, \bibinfo{author}{{Koutroumpa}, D.} et~al. (\bibinfo{year}{2010}).
\newblock \bibinfo{title}{{The Interstellar H Flow: Updated Analysis of SOHO/SWAN Data}}.
\newblock In \bibinfo{editor}{M.~{Maksimovic}}, \bibinfo{editor}{K.~{Issautier}}, \bibinfo{editor}{N.~{Meyer-Vernet}}, \bibinfo{editor}{M.~{Moncuquet}}, \& \bibinfo{editor}{F.~{Pantellini}} (Eds.), {\it \bibinfo{booktitle}{Twelfth International Solar Wind Conference}\/} (pp. \bibinfo{pages}{555--558}).
\newblock volume \bibinfo{volume}{1216} of {\it \bibinfo{series}{American Institute of Physics Conference Series}\/}.
\newblock \DOIprefix\doi{10.1063/1.3395925}. \href{http://arxiv.org/abs/1405.3474}{\tt arXiv:1405.3474}.

\bibitem[{{Lazarus} et~al.(1995){Lazarus}, {Belcher}, {Paularena}, {Richardson} \& {Steinberg}}]{lazarus1995}
\bibinfo{author}{{Lazarus}, A.~J.}, \bibinfo{author}{{Belcher}, J.~W.}, \bibinfo{author}{{Paularena}, K.~I.} et~al. (\bibinfo{year}{1995}).
\newblock \bibinfo{title}{{Recent observations of the solar wind in the outer heliosphere}}.
\newblock {\it \bibinfo{journal}{Advances in Space Research}\/},  {\it \bibinfo{volume}{16}\/}\bibinfo{issue}{(9)}, \bibinfo{pages}{77--84}. \DOIprefix\doi{10.1016/0273-1177(95)00317-8}.

\bibitem[{{Lindsay} \& {Stebbings}(2005)}]{lindsay2005}
\bibinfo{author}{{Lindsay}, B.~G.},  \& \bibinfo{author}{{Stebbings}, R.~F.} (\bibinfo{year}{2005}).
\newblock \bibinfo{title}{{Charge transfer cross sections for energetic neutral atom data analysis}}.
\newblock {\it \bibinfo{journal}{Journal of Geophysical Research (Space Physics)}\/},  {\it \bibinfo{volume}{110}\/}\bibinfo{issue}{(A12)}, \bibinfo{pages}{A12213}. \DOIprefix\doi{10.1029/2005JA011298}.

\bibitem[{{McComas} et~al.(2012{\natexlab{a}}){McComas}, {Alexashov}, {Bzowski}, {Fahr}, {Heerikhuisen}, {Izmodenov}, {Lee}, {M{\"o}bius}, {Pogorelov}, {Schwadron} \& {Zank}}]{mccomas2012a}
\bibinfo{author}{{McComas}, D.~J.}, \bibinfo{author}{{Alexashov}, D.}, \bibinfo{author}{{Bzowski}, M.} et~al. (\bibinfo{year}{2012}{\natexlab{a}}).
\newblock \bibinfo{title}{{The Heliosphere{\textquoteright}s Interstellar Interaction: No Bow Shock}}.
\newblock {\it \bibinfo{journal}{Science}\/},  {\it \bibinfo{volume}{336}\/}\bibinfo{issue}{(6086)}, \bibinfo{pages}{1291}. \DOIprefix\doi{10.1126/science.1221054}.

\bibitem[{{McComas} et~al.(2024){McComas}, {Alimaganbetov}, {Beesley}, {Bzowski}, {Funsten}, {Janzen}, {Kubiak}, {Rankin}, {Reisenfeld}, {Schwadron} \& {Szalay}}]{mccomas2024}
\bibinfo{author}{{McComas}, D.~J.}, \bibinfo{author}{{Alimaganbetov}, M.}, \bibinfo{author}{{Beesley}, L.~J.} et~al. (\bibinfo{year}{2024}).
\newblock \bibinfo{title}{{Fourteen Years of Energetic Neutral Atom Observations from IBEX}}.
\newblock {\it \bibinfo{journal}{\apjs}\/},  {\it \bibinfo{volume}{270}\/}\bibinfo{issue}{(2)}, \bibinfo{pages}{17}. \DOIprefix\doi{10.3847/1538-4365/ad0a69}.

\bibitem[{{McComas} et~al.(2009{\natexlab{a}}){McComas}, {Allegrini}, {Bochsler}, {Bzowski}, {Christian}, {Crew}, {DeMajistre}, {Fahr}, {Fichtner}, {Frisch}, {Funsten}, {Fuselier}, {Gloeckler}, {Gruntman}, {Heerikhuisen}, {Izmodenov}, {Janzen}, {Knappenberger}, {Krimigis}, {Kucharek}, {Lee}, {Livadiotis}, {Livi}, {MacDowall}, {Mitchell}, {M{\"o}bius}, {Moore}, {Pogorelov}, {Reisenfeld}, {Roelof}, {Saul}, {Schwadron}, {Valek}, {Vanderspek}, {Wurz} \& {Zank}}]{mccomas2009_science}
\bibinfo{author}{{McComas}, D.~J.}, \bibinfo{author}{{Allegrini}, F.}, \bibinfo{author}{{Bochsler}, P.} et~al. (\bibinfo{year}{2009}{\natexlab{a}}).
\newblock \bibinfo{title}{{Global Observations of the Interstellar Interaction from the Interstellar Boundary Explorer (IBEX)}}.
\newblock {\it \bibinfo{journal}{Science}\/},  {\it \bibinfo{volume}{326}\/}\bibinfo{issue}{(5955)}, \bibinfo{pages}{959}. \DOIprefix\doi{10.1126/science.1180906}.

\bibitem[{{McComas} et~al.(2009{\natexlab{b}}){McComas}, {Allegrini}, {Bochsler}, {Bzowski}, {Collier}, {Fahr}, {Fichtner}, {Frisch}, {Funsten}, {Fuselier}, {Gloeckler}, {Gruntman}, {Izmodenov}, {Knappenberger}, {Lee}, {Livi}, {Mitchell}, {M{\"o}bius}, {Moore}, {Pope}, {Reisenfeld}, {Roelof}, {Scherrer}, {Schwadron}, {Tyler}, {Wieser}, {Witte}, {Wurz} \& {Zank}}]{mccomas2009}
\bibinfo{author}{{McComas}, D.~J.}, \bibinfo{author}{{Allegrini}, F.}, \bibinfo{author}{{Bochsler}, P.} et~al. (\bibinfo{year}{2009}{\natexlab{b}}).
\newblock \bibinfo{title}{{IBEX{\textemdash}Interstellar Boundary Explorer}}.
\newblock {\it \bibinfo{journal}{\ssr}\/},  {\it \bibinfo{volume}{146}\/}\bibinfo{issue}{(1-4)}, \bibinfo{pages}{11--33}. \DOIprefix\doi{10.1007/s11214-009-9499-4}.

\bibitem[{{McComas} et~al.(2015){McComas}, {Bzowski}, {Frisch}, {Fuselier}, {Kubiak}, {Kucharek}, {Leonard}, {M{\"o}bius}, {Schwadron}, {Sok{\'o}{\l}}, {Swaczyna} \& {Witte}}]{mccomas2015}
\bibinfo{author}{{McComas}, D.~J.}, \bibinfo{author}{{Bzowski}, M.}, \bibinfo{author}{{Frisch}, P.} et~al. (\bibinfo{year}{2015}).
\newblock \bibinfo{title}{{Warmer Local Interstellar Medium: A Possible Resolution of the Ulysses-IBEX Enigma}}.
\newblock {\it \bibinfo{journal}{\apj}\/},  {\it \bibinfo{volume}{801}\/}\bibinfo{issue}{(1)}, \bibinfo{pages}{28}. \DOIprefix\doi{10.1088/0004-637X/801/1/28}.

\bibitem[{{McComas} et~al.(2018){McComas}, {Christian}, {Schwadron}, {Fox}, {Westlake}, {Allegrini}, {Baker}, {Biesecker}, {Bzowski}, {Clark}, {Cohen}, {Cohen}, {Dayeh}, {Decker}, {de Nolfo}, {Desai}, {Ebert}, {Elliott}, {Fahr}, {Frisch}, {Funsten}, {Fuselier}, {Galli}, {Galvin}, {Giacalone}, {Gkioulidou}, {Guo}, {Horanyi}, {Isenberg}, {Janzen}, {Kistler}, {Korreck}, {Kubiak}, {Kucharek}, {Larsen}, {Leske}, {Lugaz}, {Luhmann}, {Matthaeus}, {Mitchell}, {Moebius}, {Ogasawara}, {Reisenfeld}, {Richardson}, {Russell}, {Sok{\'o}{\l}}, {Spence}, {Skoug}, {Sternovsky}, {Swaczyna}, {Szalay}, {Tokumaru}, {Wiedenbeck}, {Wurz}, {Zank} \& {Zirnstein}}]{mccomas2018}
\bibinfo{author}{{McComas}, D.~J.}, \bibinfo{author}{{Christian}, E.~R.}, \bibinfo{author}{{Schwadron}, N.~A.} et~al. (\bibinfo{year}{2018}).
\newblock \bibinfo{title}{{Interstellar Mapping and Acceleration Probe (IMAP): A New NASA Mission}}.
\newblock {\it \bibinfo{journal}{\ssr}\/},  {\it \bibinfo{volume}{214}\/}\bibinfo{issue}{(8)}, \bibinfo{pages}{116}. \DOIprefix\doi{10.1007/s11214-018-0550-1}.

\bibitem[{{McComas} et~al.(2012{\natexlab{b}}){McComas}, {Dayeh}, {Allegrini}, {Bzowski}, {DeMajistre}, {Fujiki}, {Funsten}, {Fuselier}, {Gruntman}, {Janzen}, {Kubiak}, {Kucharek}, {Livadiotis}, {M{\"o}bius}, {Reisenfeld}, {Reno}, {Schwadron}, {Sok{\'o}{\l}} \& {Tokumaru}}]{mccomas2012b}
\bibinfo{author}{{McComas}, D.~J.}, \bibinfo{author}{{Dayeh}, M.~A.}, \bibinfo{author}{{Allegrini}, F.} et~al. (\bibinfo{year}{2012}{\natexlab{b}}).
\newblock \bibinfo{title}{{The First Three Years of IBEX Observations and Our Evolving Heliosphere}}.
\newblock {\it \bibinfo{journal}{\apjs}\/},  {\it \bibinfo{volume}{203}\/}\bibinfo{issue}{(1)}, \bibinfo{pages}{1}. \DOIprefix\doi{10.1088/0067-0049/203/1/1}.

\bibitem[{{McComas} et~al.(2014){McComas}, {Lewis} \& {Schwadron}}]{mccomas2014}
\bibinfo{author}{{McComas}, D.~J.}, \bibinfo{author}{{Lewis}, W.~S.},  \& \bibinfo{author}{{Schwadron}, N.~A.} (\bibinfo{year}{2014}).
\newblock \bibinfo{title}{{IBEX's Enigmatic Ribbon in the sky and its many possible sources}}.
\newblock {\it \bibinfo{journal}{Reviews of Geophysics}\/},  {\it \bibinfo{volume}{52}\/}\bibinfo{issue}{(1)}, \bibinfo{pages}{118--155}. \DOIprefix\doi{10.1002/2013RG000438}.

\bibitem[{{M{\"o}bius} et~al.(2013){M{\"o}bius}, {Liu}, {Funsten}, {Gary} \& {Winske}}]{moebius2013}
\bibinfo{author}{{M{\"o}bius}, E.}, \bibinfo{author}{{Liu}, K.}, \bibinfo{author}{{Funsten}, H.} et~al. (\bibinfo{year}{2013}).
\newblock \bibinfo{title}{{Analytic Model of the IBEX Ribbon with Neutral Solar Wind Based Ion Pickup Beyond the Heliopause}}.
\newblock {\it \bibinfo{journal}{\apj}\/},  {\it \bibinfo{volume}{766}\/}\bibinfo{issue}{(2)}, \bibinfo{pages}{129}. \DOIprefix\doi{10.1088/0004-637X/766/2/129}.

\bibitem[{{Mousavi} et~al.(2023){Mousavi}, {Liu} \& {Sadeghzadeh}}]{mousavi2023}
\bibinfo{author}{{Mousavi}, A.}, \bibinfo{author}{{Liu}, K.},  \& \bibinfo{author}{{Sadeghzadeh}, S.} (\bibinfo{year}{2023}).
\newblock \bibinfo{title}{{The Impact of Pickup Ion Thermal Spread on Pickup Ion Ring-beam-driven Instabilities and Scattering in the Outer Heliosheath}}.
\newblock {\it \bibinfo{journal}{\apj}\/},  {\it \bibinfo{volume}{958}\/}\bibinfo{issue}{(2)}, \bibinfo{pages}{151}. \DOIprefix\doi{10.3847/1538-4357/ad05be}.

\bibitem[{{Mousavi} et~al.(2025){Mousavi}, {Roytershteyn}, {Fraternale} \& {Pogorelov}}]{mousavi2025}
\bibinfo{author}{{Mousavi}, A.}, \bibinfo{author}{{Roytershteyn}, V.}, \bibinfo{author}{{Fraternale}, F.} et~al. (\bibinfo{year}{2025}).
\newblock \bibinfo{title}{{Fluctuations Driven by Multicomponent Pickup Ion Distributions in the Outer Heliosheath}}.
\newblock {\it \bibinfo{journal}{\apj}\/},  {\it \bibinfo{volume}{980}\/}\bibinfo{issue}{(1)}, \bibinfo{pages}{95}. \DOIprefix\doi{10.3847/1538-4357/adaaf4}.

\bibitem[{{Niemiec} et~al.(2016){Niemiec}, {Florinski}, {Heerikhuisen} \& {Nishikawa}}]{niemiec2016}
\bibinfo{author}{{Niemiec}, J.}, \bibinfo{author}{{Florinski}, V.}, \bibinfo{author}{{Heerikhuisen}, J.} et~al. (\bibinfo{year}{2016}).
\newblock \bibinfo{title}{{The IBEX Ribbon and the Pickup Ion Ring Stability in the Outer Heliosheath II. Monte-Carlo and Particle-in-cell Model Results}}.
\newblock {\it \bibinfo{journal}{\apj}\/},  {\it \bibinfo{volume}{826}\/}\bibinfo{issue}{(2)}, \bibinfo{pages}{198}. \DOIprefix\doi{10.3847/0004-637X/826/2/198}.

\bibitem[{{Opher} et~al.(2007){Opher}, {Stone} \& {Gombosi}}]{opher2007}
\bibinfo{author}{{Opher}, M.}, \bibinfo{author}{{Stone}, E.~C.},  \& \bibinfo{author}{{Gombosi}, T.~I.} (\bibinfo{year}{2007}).
\newblock \bibinfo{title}{{The Orientation of the Local Interstellar Magnetic Field}}.
\newblock {\it \bibinfo{journal}{Science}\/},  {\it \bibinfo{volume}{316}\/}\bibinfo{issue}{(5826)}, \bibinfo{pages}{875}. \DOIprefix\doi{10.1126/science.1139480}. \href{http://arxiv.org/abs/0705.1841}{\tt arXiv:0705.1841}.

\bibitem[{{Pogorelov} et~al.(2009){Pogorelov}, {Heerikhuisen}, {Mitchell}, {Cairns} \& {Zank}}]{pogorelov2009}
\bibinfo{author}{{Pogorelov}, N.~V.}, \bibinfo{author}{{Heerikhuisen}, J.}, \bibinfo{author}{{Mitchell}, J.~J.} et~al. (\bibinfo{year}{2009}).
\newblock \bibinfo{title}{{Heliospheric Asymmetries and 2-3 kHz Radio Emission Under Strong Interstellar Magnetic Field Conditions}}.
\newblock {\it \bibinfo{journal}{\apjl}\/},  {\it \bibinfo{volume}{695}\/}\bibinfo{issue}{(1)}, \bibinfo{pages}{L31--L34}. \DOIprefix\doi{10.1088/0004-637X/695/1/L31}.

\bibitem[{{Powell} et~al.(2024){Powell}, {Opher}, {Kornbleuth}, {Baliukin}, {Michael}, {Wood}, {Izmodenov}, {Toth} \& {Tenishev}}]{powell2024}
\bibinfo{author}{{Powell}, E.}, \bibinfo{author}{{Opher}, M.}, \bibinfo{author}{{Kornbleuth}, M.~Z.} et~al. (\bibinfo{year}{2024}).
\newblock \bibinfo{title}{{Ly{\ensuremath{\alpha}} Absorption in a ``Croissant-like'' Heliosphere}}.
\newblock {\it \bibinfo{journal}{\apj}\/},  {\it \bibinfo{volume}{961}\/}\bibinfo{issue}{(2)}, \bibinfo{pages}{235}. \DOIprefix\doi{10.3847/1538-4357/ad0cee}.

\bibitem[{{Richardson} et~al.(2022){Richardson}, {Burlaga}, {Elliott}, {Kurth}, {Liu} \& {von Steiger}}]{richardson2022}
\bibinfo{author}{{Richardson}, J.~D.}, \bibinfo{author}{{Burlaga}, L.~F.}, \bibinfo{author}{{Elliott}, H.} et~al. (\bibinfo{year}{2022}).
\newblock \bibinfo{title}{{Observations of the Outer Heliosphere, Heliosheath, and Interstellar Medium}}.
\newblock {\it \bibinfo{journal}{\ssr}\/},  {\it \bibinfo{volume}{218}\/}\bibinfo{issue}{(4)}, \bibinfo{pages}{35}. \DOIprefix\doi{10.1007/s11214-022-00899-y}.

\bibitem[{{Roytershteyn} et~al.(2019){Roytershteyn}, {Pogorelov} \& {Heerikhuisen}}]{roytershteyn2019}
\bibinfo{author}{{Roytershteyn}, V.}, \bibinfo{author}{{Pogorelov}, N.~V.},  \& \bibinfo{author}{{Heerikhuisen}, J.} (\bibinfo{year}{2019}).
\newblock \bibinfo{title}{{Pickup Ions beyond the Heliopause}}.
\newblock {\it \bibinfo{journal}{\apj}\/},  {\it \bibinfo{volume}{881}\/}\bibinfo{issue}{(1)}, \bibinfo{pages}{65}. \DOIprefix\doi{10.3847/1538-4357/ab2ad4}.

\bibitem[{{Schlickeiser}(1989)}]{schlickeiser1989}
\bibinfo{author}{{Schlickeiser}, R.} (\bibinfo{year}{1989}).
\newblock \bibinfo{title}{{Cosmic-Ray Transport and Acceleration. I. Derivation of the Kinetic Equation and Application to Cosmic Rays in Static Cold Media}}.
\newblock {\it \bibinfo{journal}{\apj}\/},  {\it \bibinfo{volume}{336}\/}, \bibinfo{pages}{243}. \DOIprefix\doi{10.1086/167009}.

\bibitem[{{Schwadron} et~al.(2009){Schwadron}, {Bzowski}, {Crew}, {Gruntman}, {Fahr}, {Fichtner}, {Frisch}, {Funsten}, {Fuselier}, {Heerikhuisen}, {Izmodenov}, {Kucharek}, {Lee}, {Livadiotis}, {McComas}, {Moebius}, {Moore}, {Mukherjee}, {Pogorelov}, {Prested}, {Reisenfeld}, {Roelof} \& {Zank}}]{schwadron2009}
\bibinfo{author}{{Schwadron}, N.~A.}, \bibinfo{author}{{Bzowski}, M.}, \bibinfo{author}{{Crew}, G.~B.} et~al. (\bibinfo{year}{2009}).
\newblock \bibinfo{title}{{Comparison of Interstellar Boundary Explorer Observations with 3D Global Heliospheric Models}}.
\newblock {\it \bibinfo{journal}{Science}\/},  {\it \bibinfo{volume}{326}\/}\bibinfo{issue}{(5955)}, \bibinfo{pages}{966}. \DOIprefix\doi{10.1126/science.1180986}.

\bibitem[{{Schwadron} \& {McComas}(2013)}]{schwadron2013}
\bibinfo{author}{{Schwadron}, N.~A.},  \& \bibinfo{author}{{McComas}, D.~J.} (\bibinfo{year}{2013}).
\newblock \bibinfo{title}{{Spatial Retention of Ions Producing the IBEX Ribbon}}.
\newblock {\it \bibinfo{journal}{\apj}\/},  {\it \bibinfo{volume}{764}\/}\bibinfo{issue}{(1)}, \bibinfo{pages}{92}. \DOIprefix\doi{10.1088/0004-637X/764/1/92}.

\bibitem[{{Schwadron} \& {McComas}(2019)}]{schwadron2019}
\bibinfo{author}{{Schwadron}, N.~A.},  \& \bibinfo{author}{{McComas}, D.~J.} (\bibinfo{year}{2019}).
\newblock \bibinfo{title}{{The Interstellar Ribbon: A Unifying Explanation}}.
\newblock {\it \bibinfo{journal}{\apj}\/},  {\it \bibinfo{volume}{887}\/}\bibinfo{issue}{(2)}, \bibinfo{pages}{247}. \DOIprefix\doi{10.3847/1538-4357/ab5b91}.

\bibitem[{{Skilling}(1971)}]{skilling1971}
\bibinfo{author}{{Skilling}, J.} (\bibinfo{year}{1971}).
\newblock \bibinfo{title}{{Cosmic Rays in the Galaxy: Convection or Diffusion?}}
\newblock {\it \bibinfo{journal}{\apj}\/},  {\it \bibinfo{volume}{170}\/}, \bibinfo{pages}{265}. \DOIprefix\doi{10.1086/151210}.

\bibitem[{{Sok{\'o}{\l}} et~al.(2022){Sok{\'o}{\l}}, {Kucharek}, {Baliukin}, {Fahr}, {Izmodenov}, {Kornbleuth}, {Mostafavi}, {Opher}, {Park}, {Pogorelov}, {Quinn}, {Smith}, {Zank} \& {Zhang}}]{sokol2022}
\bibinfo{author}{{Sok{\'o}{\l}}, J.~M.}, \bibinfo{author}{{Kucharek}, H.}, \bibinfo{author}{{Baliukin}, I.~I.} et~al. (\bibinfo{year}{2022}).
\newblock \bibinfo{title}{{Interstellar Neutrals, Pickup Ions, and Energetic Neutral Atoms Throughout the Heliosphere: Present Theory and Modeling Overview}}.
\newblock {\it \bibinfo{journal}{\ssr}\/},  {\it \bibinfo{volume}{218}\/}\bibinfo{issue}{(3)}, \bibinfo{pages}{18}. \DOIprefix\doi{10.1007/s11214-022-00883-6}.

\bibitem[{{Sok{\'o}{\l}} et~al.(2020){Sok{\'o}{\l}}, {McComas}, {Bzowski} \& {Tokumaru}}]{sokol_etal:20a}
\bibinfo{author}{{Sok{\'o}{\l}}, J.~M.}, \bibinfo{author}{{McComas}, D.~J.}, \bibinfo{author}{{Bzowski}, M.} et~al. (\bibinfo{year}{2020}).
\newblock \bibinfo{title}{{Sun-Heliosphere Observation-based Ionization Rates Model}}.
\newblock {\it \bibinfo{journal}{\apj}\/},  {\it \bibinfo{volume}{897}\/}\bibinfo{issue}{(2)}, \bibinfo{pages}{179}. \DOIprefix\doi{10.3847/1538-4357/ab99a4}. \href{http://arxiv.org/abs/2003.09292}{\tt arXiv:2003.09292}.

\bibitem[{{Stone} et~al.(2008){Stone}, {Cummings}, {McDonald}, {Heikkila}, {Lal} \& {Webber}}]{stone2008}
\bibinfo{author}{{Stone}, E.~C.}, \bibinfo{author}{{Cummings}, A.~C.}, \bibinfo{author}{{McDonald}, F.~B.} et~al. (\bibinfo{year}{2008}).
\newblock \bibinfo{title}{{An asymmetric solar wind termination shock}}.
\newblock {\it \bibinfo{journal}{\nat}\/},  {\it \bibinfo{volume}{454}\/}\bibinfo{issue}{(7200)}, \bibinfo{pages}{71--74}. \DOIprefix\doi{10.1038/nature07022}.

\bibitem[{{Swaczyna} et~al.(2016){Swaczyna}, {Bzowski}, {Christian}, {Funsten}, {McComas} \& {Schwadron}}]{swaczyna2016}
\bibinfo{author}{{Swaczyna}, P.}, \bibinfo{author}{{Bzowski}, M.}, \bibinfo{author}{{Christian}, E.~R.} et~al. (\bibinfo{year}{2016}).
\newblock \bibinfo{title}{{Distance to the IBEX Ribbon Source Inferred from Parallax}}.
\newblock {\it \bibinfo{journal}{\apj}\/},  {\it \bibinfo{volume}{823}\/}\bibinfo{issue}{(2)}, \bibinfo{pages}{119}. \DOIprefix\doi{10.3847/0004-637X/823/2/119}. \href{http://arxiv.org/abs/1603.09134}{\tt arXiv:1603.09134}.

\bibitem[{{Swaczyna} et~al.(2020){Swaczyna}, {McComas}, {Zirnstein}, {Sok{\'o}{\l}}, {Elliott}, {Bzowski}, {Kubiak}, {Richardson}, {Kowalska-Leszczynska}, {Stern}, {Weaver}, {Olkin}, {Singer} \& {Spencer}}]{swaczyna_etal:20a}
\bibinfo{author}{{Swaczyna}, P.}, \bibinfo{author}{{McComas}, D.~J.}, \bibinfo{author}{{Zirnstein}, E.~J.} et~al. (\bibinfo{year}{2020}).
\newblock \bibinfo{title}{{Density of Neutral Hydrogen in the Sun's Interstellar Neighborhood}}.
\newblock {\it \bibinfo{journal}{\apj}\/},  {\it \bibinfo{volume}{903}\/}\bibinfo{issue}{(1)}, \bibinfo{pages}{48}. \DOIprefix\doi{10.3847/1538-4357/abb80a}.

\bibitem[{{Williams} et~al.(1995){Williams}, {Zank} \& {Matthaeus}}]{williams1995}
\bibinfo{author}{{Williams}, L.~L.}, \bibinfo{author}{{Zank}, G.~P.},  \& \bibinfo{author}{{Matthaeus}, W.~H.} (\bibinfo{year}{1995}).
\newblock \bibinfo{title}{{Dissipation of pickup-induced waves: A solar wind temperature increase in the outer heliosphere?}}
\newblock {\it \bibinfo{journal}{\jgr}\/},  {\it \bibinfo{volume}{100}\/}\bibinfo{issue}{(A9)}, \bibinfo{pages}{17059--17068}. \DOIprefix\doi{10.1029/95JA01261}.

\bibitem[{{Witte}(2004)}]{witte2004}
\bibinfo{author}{{Witte}, M.} (\bibinfo{year}{2004}).
\newblock \bibinfo{title}{{Kinetic parameters of interstellar neutral helium. Review of results obtained during one solar cycle with the Ulysses/GAS-instrument}}.
\newblock {\it \bibinfo{journal}{\aap}\/},  {\it \bibinfo{volume}{426}\/}, \bibinfo{pages}{835--844}. \DOIprefix\doi{10.1051/0004-6361:20035956}.

\bibitem[{{Zank} et~al.(2013){Zank}, {Heerikhuisen}, {Wood}, {Pogorelov}, {Zirnstein} \& {McComas}}]{zank2013}
\bibinfo{author}{{Zank}, G.~P.}, \bibinfo{author}{{Heerikhuisen}, J.}, \bibinfo{author}{{Wood}, B.~E.} et~al. (\bibinfo{year}{2013}).
\newblock \bibinfo{title}{{Heliospheric Structure: The Bow Wave and the Hydrogen Wall}}.
\newblock {\it \bibinfo{journal}{\apj}\/},  {\it \bibinfo{volume}{763}\/}\bibinfo{issue}{(1)}, \bibinfo{pages}{20}. \DOIprefix\doi{10.1088/0004-637X/763/1/20}.

\bibitem[{{Zirnstein} et~al.(2016{\natexlab{a}}){Zirnstein}, {Funsten}, {Heerikhuisen} \& {McComas}}]{zirnstein2016a}
\bibinfo{author}{{Zirnstein}, E.~J.}, \bibinfo{author}{{Funsten}, H.~O.}, \bibinfo{author}{{Heerikhuisen}, J.} et~al. (\bibinfo{year}{2016}{\natexlab{a}}).
\newblock \bibinfo{title}{{Effects of solar wind speed on the secondary energetic neutral source of the Interstellar Boundary Explorer ribbon}}.
\newblock {\it \bibinfo{journal}{\aap}\/},  {\it \bibinfo{volume}{586}\/}, \bibinfo{pages}{A31}. \DOIprefix\doi{10.1051/0004-6361/201527437}.

\bibitem[{{Zirnstein} et~al.(2016{\natexlab{b}}){Zirnstein}, {Funsten}, {Heerikhuisen}, {McComas}, {Schwadron} \& {Zank}}]{zirnstein2016b}
\bibinfo{author}{{Zirnstein}, E.~J.}, \bibinfo{author}{{Funsten}, H.~O.}, \bibinfo{author}{{Heerikhuisen}, J.} et~al. (\bibinfo{year}{2016}{\natexlab{b}}).
\newblock \bibinfo{title}{{Geometry and Characteristics of the Heliosheath Revealed in the First Five Years of Interstellar Boundary Explorer Observations}}.
\newblock {\it \bibinfo{journal}{\apj}\/},  {\it \bibinfo{volume}{826}\/}\bibinfo{issue}{(1)}, \bibinfo{pages}{58}. \DOIprefix\doi{10.3847/0004-637X/826/1/58}.

\bibitem[{{Zirnstein} et~al.(2018){Zirnstein}, {Heerikhuisen} \& {Dayeh}}]{zirnstein2018}
\bibinfo{author}{{Zirnstein}, E.~J.}, \bibinfo{author}{{Heerikhuisen}, J.},  \& \bibinfo{author}{{Dayeh}, M.~A.} (\bibinfo{year}{2018}).
\newblock \bibinfo{title}{{The Role of Pickup Ion Dynamics Outside of the Heliopause in the Limit of Weak Pitch Angle Scattering: Implications for the Source of the IBEX Ribbon}}.
\newblock {\it \bibinfo{journal}{\apj}\/},  {\it \bibinfo{volume}{855}\/}\bibinfo{issue}{(1)}, \bibinfo{pages}{30}. \DOIprefix\doi{10.3847/1538-4357/aaaf6d}.

\bibitem[{{Zirnstein} et~al.(2015){Zirnstein}, {Heerikhuisen} \& {McComas}}]{zirnstein2015b}
\bibinfo{author}{{Zirnstein}, E.~J.}, \bibinfo{author}{{Heerikhuisen}, J.},  \& \bibinfo{author}{{McComas}, D.~J.} (\bibinfo{year}{2015}).
\newblock \bibinfo{title}{{Structure of the Interstellar Boundary Explorer Ribbon from Secondary Charge-exchange at the Solar-Interstellar Interface}}.
\newblock {\it \bibinfo{journal}{\apjl}\/},  {\it \bibinfo{volume}{804}\/}\bibinfo{issue}{(1)}, \bibinfo{pages}{L22}. \DOIprefix\doi{10.1088/2041-8205/804/1/L22}.

\bibitem[{{Zirnstein} et~al.(2013){Zirnstein}, {Heerikhuisen}, {McComas} \& {Schwadron}}]{zirnstein2013}
\bibinfo{author}{{Zirnstein}, E.~J.}, \bibinfo{author}{{Heerikhuisen}, J.}, \bibinfo{author}{{McComas}, D.~J.} et~al. (\bibinfo{year}{2013}).
\newblock \bibinfo{title}{{Simulating the Compton-Getting Effect for Hydrogen Flux Measurements: Implications for IBEX-Hi and -Lo Observations}}.
\newblock {\it \bibinfo{journal}{\apj}\/},  {\it \bibinfo{volume}{778}\/}\bibinfo{issue}{(2)}, \bibinfo{pages}{112}. \DOIprefix\doi{10.1088/0004-637X/778/2/112}.

\bibitem[{{Zirnstein} et~al.(2019{\natexlab{a}}){Zirnstein}, {McComas}, {Schwadron}, {Dayeh}, {Heerikhuisen} \& {Swaczyna}}]{zirnstein2019}
\bibinfo{author}{{Zirnstein}, E.~J.}, \bibinfo{author}{{McComas}, D.~J.}, \bibinfo{author}{{Schwadron}, N.~A.} et~al. (\bibinfo{year}{2019}{\natexlab{a}}).
\newblock \bibinfo{title}{{Strong Scattering of {\ensuremath{\sim}}keV Pickup Ions in the Local Interstellar Magnetic Field Draped around Our Heliosphere: Implications for the IBEX Ribbon's Source and IMAP}}.
\newblock {\it \bibinfo{journal}{\apj}\/},  {\it \bibinfo{volume}{876}\/}\bibinfo{issue}{(2)}, \bibinfo{pages}{92}. \DOIprefix\doi{10.3847/1538-4357/ab15d6}.

\bibitem[{{Zirnstein} et~al.(2023){Zirnstein}, {Swaczyna}, {Dayeh} \& {Heerikhuisen}}]{zirnstein2023}
\bibinfo{author}{{Zirnstein}, E.~J.}, \bibinfo{author}{{Swaczyna}, P.}, \bibinfo{author}{{Dayeh}, M.~A.} et~al. (\bibinfo{year}{2023}).
\newblock \bibinfo{title}{{Constraints on the IBEX Ribbon's Origin from Its Evolution over a Solar Cycle}}.
\newblock {\it \bibinfo{journal}{\apj}\/},  {\it \bibinfo{volume}{949}\/}\bibinfo{issue}{(2)}, \bibinfo{pages}{45}. \DOIprefix\doi{10.3847/1538-4357/acc577}.

\bibitem[{{Zirnstein} et~al.(2019{\natexlab{b}}){Zirnstein}, {Swaczyna}, {McComas} \& {Heerikhuisen}}]{zirnstein2019_parallax}
\bibinfo{author}{{Zirnstein}, E.~J.}, \bibinfo{author}{{Swaczyna}, P.}, \bibinfo{author}{{McComas}, D.~J.} et~al. (\bibinfo{year}{2019}{\natexlab{b}}).
\newblock \bibinfo{title}{{Parallax of the IBEX Ribbon Indicates a Spatially Retained Source}}.
\newblock {\it \bibinfo{journal}{\apj}\/},  {\it \bibinfo{volume}{879}\/}\bibinfo{issue}{(2)}, \bibinfo{pages}{106}. \DOIprefix\doi{10.3847/1538-4357/ab2633}.

\end{thebibliography}



\appendix

\section{Spatial distribution of the ribbon flux production rate}\label{app:spatial_sources}

The ribbon ENAs can be observed only from specific directions in the sky. These directions depend on the distribution of primary ENA sources, the position of the observer, and the energy of registered secondary ENAs. In this section, we perform a qualitative study of the spatial distribution of the ribbon production rate outside the heliopause and analyze its dependence on the aforementioned parameters. 

\subsection{Derivation of the geometrical condition} \label{sec:geometric_condition}

First, we derive the analytical condition that defines the surface, where the ribbon production rate is maximal. Let us consider a point source of atoms at $\mathbf{r}_{\rm src}$ with velocity $v$ and assume that a primary ENA from this point source charge exchanges at some arbitrary point $\mathbf{r}_1$ outside the heliopause and becomes a pickup proton. The pickup proton at this point has a pitch-angle cosine 
\begin{equation}
\mu_1(\mathbf{r}_1, \mathbf{r}_{\rm src}, v) = \frac{\mathbf{w}_1 \cdot \mathbf{b}}{w_1},
\end{equation}
where $\mathbf{w}_1 = \mathbf{v}_1 - \mathbf{U}$ and $\mathbf{v}_1 = v_1 \mathbf{e}_1$ are the velocities of the primary ENA in the plasma and inertial reference frame, and $\mathbf{e}_1 = (\mathbf{r}_1 - \mathbf{r}_{\rm src}) / |\mathbf{r}_1 - \mathbf{r}_{\rm src}|$ is the direction of its motion from the point source at $\mathbf{r} = \mathbf{r}_{\rm src}$.

The observer at $\mathbf{r}_{\rm obs}$ registers a secondary ENA that originates at the arbitrary point $\mathbf{r}_2$ with velocity $v$ if the parent pickup proton has the pitch angle cosine
\begin{equation}
\mu_2(\mathbf{r}_2, \mathbf{r}_{\rm obs}, v) = \frac{\mathbf{w}_2 \cdot \mathbf{b}}{w_2},
\end{equation}
where $\mathbf{w}_2 = \mathbf{v}_2 - \mathbf{U}$ and $\mathbf{v}_2 = v_2 \mathbf{e}_2$ are the velocities of the secondary ENA in the plasma and inertial reference frame, and $\mathbf{e}_2 = (\mathbf{r}_{\rm obs} - \mathbf{r}_2) / |\mathbf{r}_{\rm obs} - \mathbf{r}_2|$ is the direction of its motion towards the observer at $\mathbf{r} = \mathbf{r}_{\rm obs}$.

Now, let us assume that pickup protons do not change position and energy during their lifespan, i.e., the time between their birth (charge exchange of the primary ENAs) and subsequent neutralization. Note that we do not use this assumption in the numerical simulations. In our model, the position, velocity, and pitch-angle of PUIs change according to equations (\ref{eq:characteristic}). The purpose of introducing this assumption is to enable an analytical study of the spatial distribution of the production rate in a zero-order approximation. Even though the characteristic time of the pickup proton neutralization in the outer heliosheath is relatively long ($\sim$few years for energies under consideration), this assumption is roughly satisfied in the regions of particular interest since $\mu \approx 0$ there. So, we assume that $\mathbf{r}_1 = \mathbf{r}_2 = \mathbf{r}$ and $v_1 = v_2 = v$. In that case, the following condition defines points $\mathbf{r}$ outside the heliopause where the re-neutralization of ENAs starting from $\mathbf{r}_{\rm src}$ happens at the appropriate angle, so the observer at $\mathbf{r}_{\rm obs}$ can register them:
\begin{equation}
\Delta \mu(\mathbf{r}, \mathbf{r}_{\rm src}, \mathbf{r}_{\rm obs}, v) = \mu_1(\mathbf{r}, \mathbf{r}_{\rm src}, v) - \mu_2(\mathbf{r}, \mathbf{r}_{\rm obs}, v) = 0.
\label{eq:reflection_condition_general}
\end{equation}
We note that this condition defines a surface in space and depends on the velocity $v$ of primary/observed ENAs. However, $U \ll v$ and, therefore, $w_1 \approx w_2$ for ENAs with energies of a few keV (IBEX-Hi energy range), so the condition (\ref{eq:reflection_condition_general}) can be simplified to
\begin{equation}
\mathbf{e}_1(\mathbf{r}, \mathbf{r}_{\rm src}) \cdot \mathbf{b}(\mathbf{r}) = \mathbf{e}_2(\mathbf{r}, \mathbf{r}_{\rm obs}) \cdot \mathbf{b}(\mathbf{r}).
\label{eq:reflection_condition_simplified}
\end{equation}
It can be interpreted as ``the law of reflection'': the angle of incidence (of primary ENAs to the magnetic field line) should be equal to the angle of reflection (of secondary ENAs from the magnetic field line). Therefore, the ribbon observed by IBEX, in some sense, can be considered as an optical effect.

In the simplest case of a point source of primary ENAs at the Sun (analog of the NSW population) and an observer near the Sun (analog of the IBEX spacecraft), i.e., $\mathbf{r}_{\rm src} = \mathbf{r}_{\rm obs} =\mathbf{0}$, we have $\mathbf{e}_1 = -\mathbf{e}_2 = \mathbf{e}_{\rm r}$ (radial direction) and get the following commonly used condition:
\begin{equation}
\mathbf{r} \cdot \mathbf{B} = 0.
\label{eq:reflection_condition}
\end{equation}

\subsection{Analysis of the spatial distribution}

\begin{figure*}
\includegraphics[width=\textwidth]{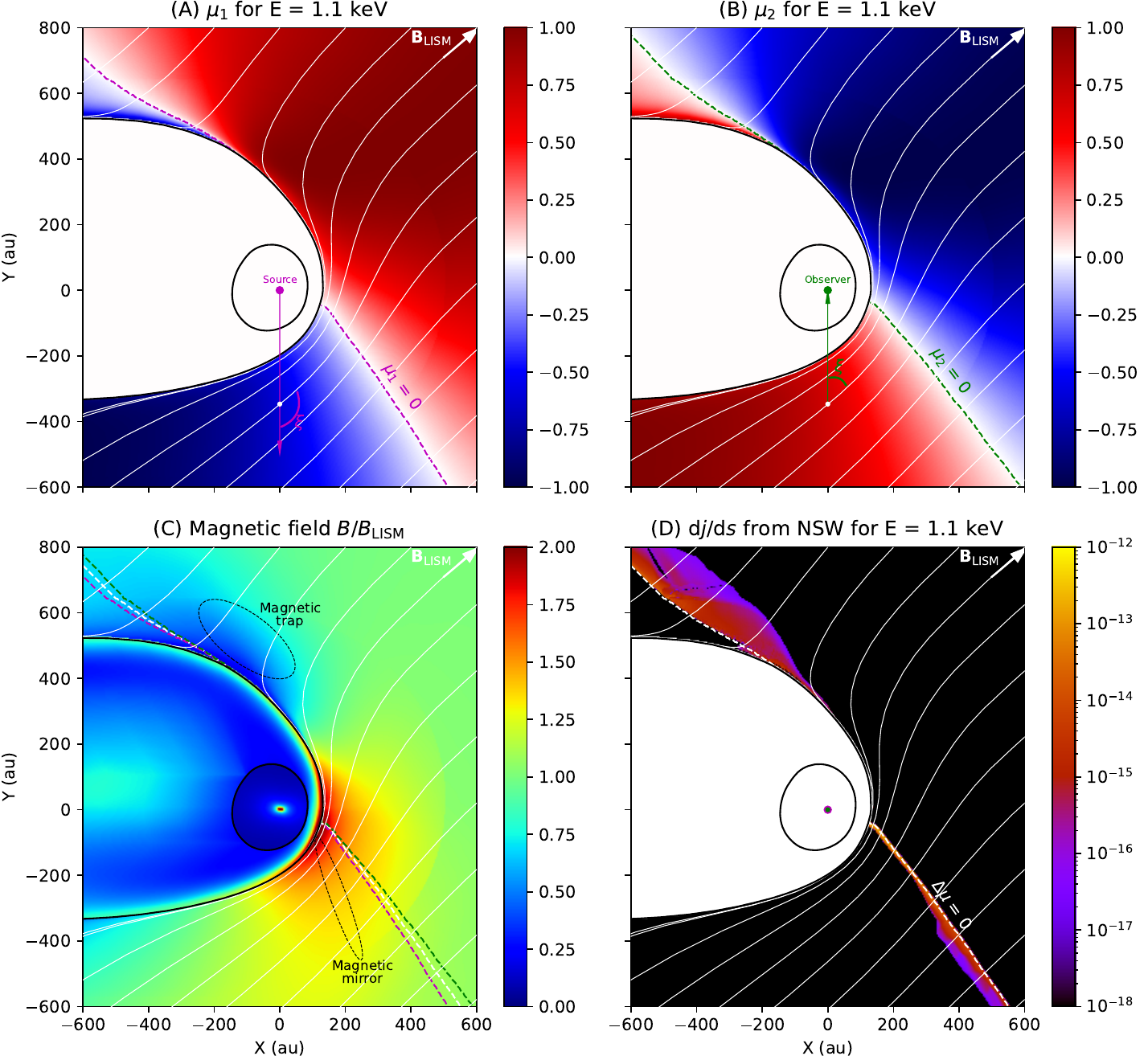}
\caption{
Spatial distribution (in the BV-plane) of the pitch-angle ($\xi$) cosines $\mu_1$ and $\mu_2$ (panels A and B), normalized magnetic field (panel C), and the ribbon production rate $\mathrm{d}j/\mathrm{d}s$ (cm$^3$ sr s keV)$^{-1}$ from the NSW component (panel D). $\mu_1$ and $\mu_2$ are calculated assuming both the point source of primary ENAs and observer at the Sun. Pitch-angle cosines and production rate are calculated for the specific energy $E$ = 1.1 keV. Black and white lines show the TS/HP boundaries and magnetic field lines, respectively. Magenta and green dashed lines show the points where $\mu_1$ and $\mu_2$ are zero, respectively. The white dashed line shows points where $\mu_1 = \mu_2$.
}
\label{fig:RIB_dynamics}
\end{figure*}

Figure \ref{fig:RIB_dynamics} shows the spatial distribution (in the BV-plane) of the pitch-angle cosines $\mu_1$ and $\mu_2$ (panels A and B), normalized magnetic field (panel C), and the ribbon production rate $\mathrm{d}j/\mathrm{d}s$ from the NSW component (panel D). $\mu_1$ and $\mu_2$ are calculated assuming both the point source of primary ENAs and observer at the Sun. The pitch-angle cosines and production rate are calculated for a specific velocity $v = \sqrt{2E / m_{\rm H}}$ that corresponds to the energy $E$ = 1.1 keV  (central energy of ESA3 of IBEX-Hi). We note that even though $\mu_1$ and $\mu_2$ are dependent on the direction of the interstellar magnetic field, all the following conclusions are held (except for the sign) if the opposite field orientation is assumed. 

As can be seen from Figure \ref{fig:RIB_dynamics}, $\mu_1$ and $\mu_2$ are somewhat inverted in sign. The only region where $\mu_1$ and $\mu_2$ have the same sign (both positive) is between the magenta and green lines (both lines are shown in panel C), which do not coincide because $U \neq 0$. 

So, this region is preferential for the ribbon production, with the best conditions along the white dashed line, where $\Delta \mu = \mu_1 - \mu_2$ = 0 (equivalent to $\mathbf{r} \cdot \mathbf{B} = 0$). As can be seen in panel (D), the maximum production rate is close to this line, so our model simulations are in correspondence with the analytic equation (\ref{eq:reflection_condition_general}). 

Panel (D) of Figure \ref{fig:RIB_dynamics} shows that there are two regions adjacent to the dashed white lines, where the production rate is also relatively high. One region has a positive $Y$-coordinate and the other has a negative one. The non-zero values of the production rate in these regions can be understood if we remember that the pitch-angle actually changes along the pickup proton trajectory (contrary to the simplified assumption made during the derivation of the geometrical condition in \ref{sec:geometric_condition}).

In the region at the top, to the right of the white dashed line, pickup protons initially have a small positive pitch angle (see panel A), so they move along the magnetic field line (with increasing $X$ and $Y$ coordinates). However, the magnetic field is increasing in this direction, which means that $\mathrm{div}\mathbf{b} > 0$ and, according to equation \ref{eq:dmu_dt_divb}), the pitch angle derivative is negative. At some point, the pitch angle becomes negative, so pickup protons change their direction and move along the magnetic field line in the opposite direction (magnetic mirroring effect). A negative pitch angle in this region is needed for their daughter's secondary ENAs to be observed (see panel B). We also note that close to the heliopause, the top region of enhanced ribbon production is adjacent to the so-called magnetic trap, where the magnetic field is low \citep[see, e.g.,][]{florinski2024}. 

In the region at the bottom, to the left of the white dashed line, the situation is identical. Pickup protons with initially small negative pitch angle (see panel A) move along the magnetic field line in the opposite direction, and since the magnetic field is increasing, they bounce back from the magnetic mirror and change their direction of motion (note that the maximum of the magnetic field is shifted to the left of $\Delta \mu = 0$). As a result, the particles acquire a positive pitch angle necessary for secondary ENAs to be observed near the Sun (see panel B). We note that the production rate in the bottom region is generally higher than in the top region since it is closer to the source of primary ENAs (supersonic solar wind region for NSW atoms), whose number density after the TS is falling rapidly with distance from the Sun. 

Therefore, we can conclude that the regions where $\mu_1$ and $\mathrm{d}\mu/\mathrm{d}t$ have different signs are also preferential for the ribbon ENAs production. However, we note that pickup protons that change the direction of their motion spend a relatively long time before they acquire the necessary pitch angle. Consequently, their abundance is rapidly decreasing over time due to charge exchange extinction, which results in relatively low production rates in the corresponding regions (compared to points at the principal line $\Delta \mu = 0$).

\subsection{Dependence on the positions of observer and primary source}

\begin{figure*}
\includegraphics[width=\textwidth]{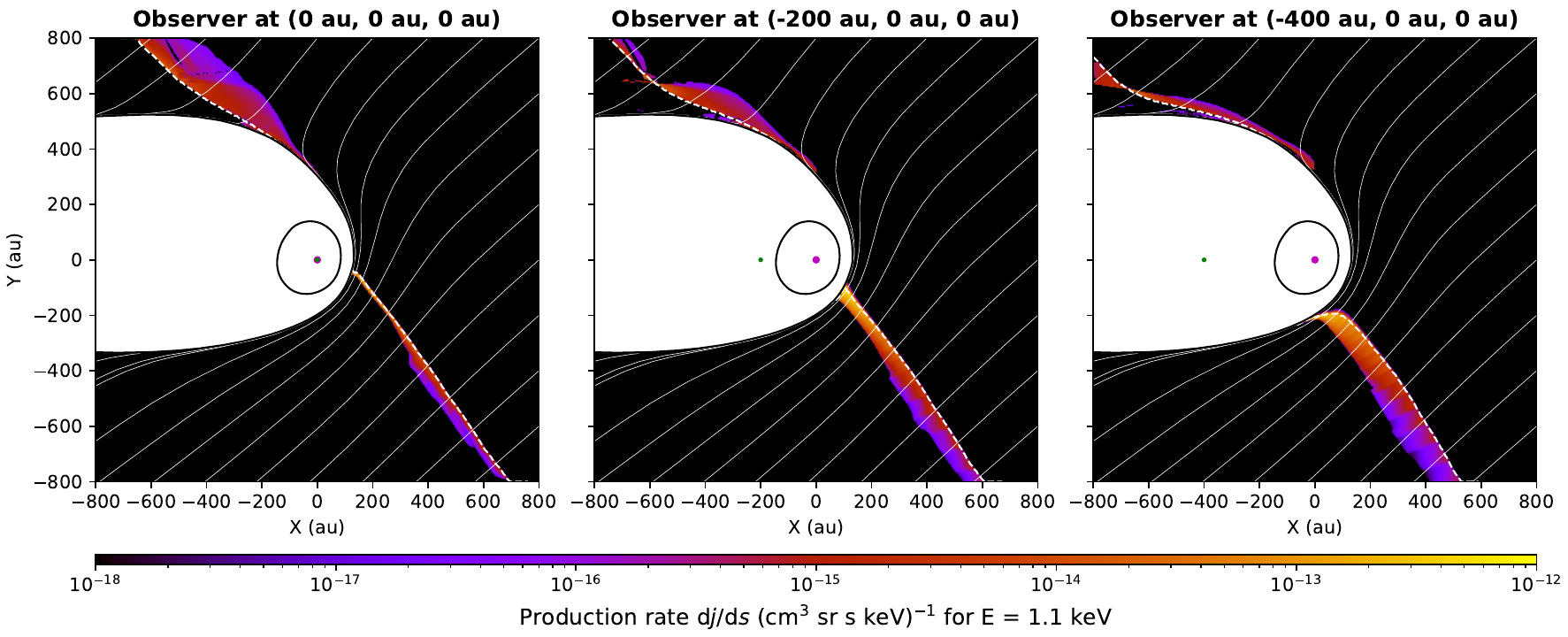}
\caption{
Spatial distribution of the ribbon production rate from the NSW primary ENAs for three positions of the observer (shown with green dots): at 0 au, 200 au, and 400 au in the downwind direction. The production rate is calculated for the energy $E$ = 1.1 keV of observed ENAs. Black and white lines show the TS/HP boundaries and magnetic field lines, respectively. The white dashed line shows points where $\Delta \mu = 0$.
}
\label{fig:RIB_3observers}
\end{figure*}

Figure \ref{fig:RIB_3observers} shows the spatial distribution of the ribbon production rate from the NSW primary ENAs for three observer positions: at 0 au, 200 au, and 400 au in the downwind direction. The production rate is calculated for the energy $E$ = 1.1 keV of observed ENAs. It is clearly seen in this figure that the spatial distribution of the ribbon production rate is dependent on the observer's position. As seen in Figure \ref{fig:RIB_3observers}, the white dashed lines, which show points where the condition $\Delta \mu = 0$ is satisfied, shift to the left in correspondence with the observer's position. The model calculations suggest that these analytical lines approximate the points of non-zero production rate well.

We also note a symmetry in the geometric condition (\ref{eq:reflection_condition_simplified}) between the ``incident'' primary and ``reflected'' secondary ENAs: if we swap the source and observer positions, the condition is still satisfied. Therefore, Figure \ref{fig:RIB_3observers} also helps to understand the distribution of the ribbon production rate from other populations of primary ENAs. In the case of NPIs, which are produced throughout SSW region, one can expect a broader region of non-zero production rate outside the heliopause, as seen by the observer at the Sun (compared to the NSW). This broadening should be even more prominent for the component of primary ENAs from the IHS, whose source region is even wider (between the TS and HP). The calculations of the ribbon production rate for three components of primary ENAs, which support these results, are shown in Figure \ref{fig:BV40_sources}.

It is important to note that all the conclusions above are made using the basic assumption of no pitch-angle scattering to describe the pickup proton dynamics outside the heliopause. As seen from the modeling results, in this case, the ribbon production rate outside the heliopause strongly depends on the observer position. This means that the source of ribbon ENAs is not spatially confined, and the estimation of the distance to it using the parallax effect should be complicated.

Nevertheless, \citet{swaczyna2016} concluded that the observation of the ribbon parallax is feasible and estimated the distance to the source as 140$^{+84}_{-38}$ au based on the IBEX-Hi data at high ecliptic latitudes obtained from the opposite sides of the Earth position around the Sun (baseline of $\sim$2 au). Later, \citet{zirnstein2019_parallax} performed simulations of the ribbon parallax with different assumptions on the pitch-angle efficiency outside the heliopause. They showed that the modeling results with the spatial retention mechanism of the ribbon production agree with the apparent parallax of the source region, and the authors concluded that the IBEX ribbon source is spatially confined. However, their modeled fluxes required a significant scaling (by a factor of $\sim$4) to fit the data, which is the major issue of the spatial retention model.

\end{document}